# PECR : A formal system based on computability logic

*G. Pantelis*

March 22, 2024

**Abstract:** *PECR is a formal system designed to explore the properties of computability of programs on a real-world computer. As such PECR incorporates the finite resources of the machine upon which a program is to be executed. The main features of the formal system will be presented and its practical applications will be discussed. Of particular interest is the implementation of the formal system to the exploration of the laws of nature that lead to rigorous constructions of computer models of real-world phenomena.*

## Table of Contents

# 1 Introduction

## 1.1 Background

Mathematics is credited for having acquired an enormous amount of knowledge of abstract objects of the platonic universe. While much of the effort in this area has largely been motivated by purely academic interest, scientist have found in these objects a way to express their understanding of real-world phenomena.

While theories of real-world phenomena constructed in this way may appear to be elegant and satisfying the abstractions embedded in them introduce complications that are largely overlooked. In particular, theories expressed in the language of continuous mathematics, often in the form of second-order partial differential equations, require considerable effort to translate them into a language of computation. Despite the enormous advances in the subject of numerical analysis, a complete rigorous validation of a computer model as an approximation of its theoretical template is often not achievable. This is an unfortunate state of affairs given that scientific research is becoming increasingly reliant on computer models.

This issue has largely been brushed aside as being due to the limitations of computers and their inability to recognize the abstract objects of mathematics. What is rarely considered is the possibility that the problem may be the other way around. Namely, it is the theoretical model that is expressed in a language that is incompatible with real-world applications.

Adding weight to this possibility is the fact that in many applications, particularly in the area of fluid dynamics and more generally hydrodynamics, the existence of solutions of the mathematical formulations have yet to be established. Notable are the Navier-Stokes equations for incompressible fluid flow where, despite considerable efforts, a proof of the existence of solutions under typical initial and boundary conditions remains elusive. For some this comes as no surprise given the compelling arguments supporting the claim that such solutions cannot exist [1].

Although somewhat tempered in modern times there still remains an attitude, held since antiquity, that the highest form of knowledge is to be acquired through explorations of the platonic universe and that the real-world occupies the lower status of just being an approximation of it. Since the advent of computers experiments in computation have revealed that there is a wealth of knowledge to be acquired under the hypothesis of a computable universe, the richness of which is becoming increasingly apparent.

Before any attempt is made to delve deeper into this topic it is worthwhile to briefly reexamine the very foundations of mathematics itself. In the early days of the development of set theory a number of paradoxes were discovered, often involving infinities. These were of great concern, but the first major shock to mathematics and formal systems in general came in the form of Godel's incompleteness theorem. Despite this most pure mathematicians have carried on as usual in their special subject area of interest adhering to the formal spirit of the axiomatic method. In more recent times some mathematicians and scientist have raised serious questions of this practice and have even proposed that we abandon the axiomatic method altogether in favor of a more constructive approach by way of experimental mathematics.

However the idea of moving towards experimental mathematics does not resolve the issue for those who insist that computer models should meet some higher standards of rigor. In this respect we are forced to look elsewhere to find a resolution.

Over recent decades has emerged an increasing acceptance among physicists that information is fundamental to observable objects and their dynamic behavior. The early proposals of digital physics are now becoming a serious topic of discussion in the wider scientific community [2]-[5]. The advantage of any phenomena that can be described by a completely discrete formulation is that its exportation into a language of computation and programs becomes far more natural. In this way the formulation of the theory is embedded, in its entirety, within the computer model bypassing the need to introduce any kind of approximation. Consequently, the core of the actual program code of the computer model can be regarded as essentially taking on the role of the defining expression of the laws governing the particular application under consideration. This supports a paradigm shift from the traditional convention of expressing the laws of nature by way of mathematical equations to one of programs.

In the current paradigm of the sciences a theory is regarded as rigorously defined if it can be constructed through the conventional language of mathematics. This includes the computer sciences where explorations of computability are carried out using the standard style of analysis employed by pure mathematicians. These often involve the introduction of abstractions of computers by way of Turing machines as well as abstractions of functional programming languages through the lambda calculus [6].

Our objective here is to construct a formal language whereby explorations of computability can be conducted entirely by real-world machine computation without the need to draw upon such abstractions. To this end the abstractions embedded in contemporary proof theory (see for example [7]-[8]) along with the lambda calculus become inappropriate and more primitive tools are required.

While acknowledging the merits of the above mentioned arguments on the shortcomings of the axiomatic method we shall take a less extreme approach and recognize that formal systems have an important role to play alongside experimental computation. The interplay between formal systems and experimental computation has been discussed in the final chapter of [9].

The material presented throughout the course of these notes has been influenced, at least in parts, by the works of a number of pioneers in the sciences and mathematics. Notable among these are the works of Stephen Wolfram who is one of the earliest figures to argue in favor of expressing real-world phenomena in a language of programs rather than mathematical equations [10].

Other notable figures include the mathematician Doron Zeilberger who rejects the existence of real numbers. His ideas on ultrafinitism and his attacks on the axiomatic method have made him a somewhat controversial figure. He is one of the most outspoken proponents of abandoning the practice of mathematical proof in favor of experimental mathematics [11]-[12].

There is also the works of Gregory Chaitin [13]-[22] who has developed an incompleteness theorem based on the complexity of programs that generate proofs of theorems from formal axiomatic systems of mathematics. This has added further weight to the notion that mathematics is falling short of the absolute certainty that it was once hoped to be.

Another mathematician who also raises questions of contemporary mathematics is Norman Wildberger, who is credited for developing the subject of rational trigonometry [23]-[25]. Norman Wildberger has made serious attempts to revise much of contemporary mathematics by excluding notions of infinite sets and the real number system.

It would require much more space to document all of the names that have made significant contributions to this topic. It suffices to say that there appears to be a growing movement where concerns of contemporary mathematics are being raised with favor given to theories that are computationally valid.

## 1.2  Real-world machine computation

In the formal system PECR statements are expressed as strings representing well defined functional programs. The aim is to establish the computability of a program made up of an ordered list of atomic functional programs subject to a collection of inference and construction rules. PECR is based on *computability logic* (CoL) that is implemented entirely by machine computation. In this way the process of rigorous program construction becomes synonymous with proof based on CoL.

The notion of computability logic was first introduced in 2003 by Giorgi Japaridze [26] who described it as a mathematical framework for redeveloping logic as a systematic formal theory of computability, as opposed to classical logic which is a formal theory of truth. The formal system PECR is a version of computability logic that differs from that developed in [26]. In [26] computability logic is formalized as a game played by a machine against its environment whereas the rules of CoL in PECR emerge from a more direct approach that is largely dictated by program structures.

Another important feature of PECR is that it is not presented as a theory that is to be explored by conventional mathematical analysis but rather has been designed purely as a computational tool that is to be implemented directly onto a machine platform. The intention being that explorations of any application that can be expressed in the language of programs is to be carried out interactively with the machine itself. *Essential to these explorations is the incorporation of the properties of the limited resources of the machine that define the environment within which the application is to be implemented.*

The principal inference rule of PECR is called the *program extension rule*. This is accompanied by additional construction rules that, in the most part, define the properties of program operations and type assignments. The inference and construction rules of PECR are expressed in the form of *irreducible extended programs*. Each statement of these rules is represented by a higher order program. Higher order programs are functional programs whose I/O list elements can be assigned the values of programs.

When working in the context of the formal system PECR we will often use the word *application* instead of the word *theory* because the latter has suggestions of abstraction. Each application of PECR comes with its own atomic programs and axioms. These application specific axioms are distinguished from the axioms of the underlying formal system, PECR, that come in the form of general inference and construction rules.

A major objective in the construction of our formal system is to design a language that is simple to use without compromising its power of analysis. As such the language is easy to learn by those with little background in the computer sciences. It should be of interest to those who can accept that problems of real-world applications can be posed as a list of functional programs subject to validation based on computability. Another important motivation behind the choice of the syntax employed in the language is the anticipation of efficiently coupling the formal system with machine learning methods that lead to increased automation.

The first 5 chapters describe the foundations of the formal language PECR. Chapters 6-7 introduce rules for disjunctions and conjunctions. These can be regarded as supplemental to the main core axioms of PECR. Chapters 8-9 discuss some specific applications of the formal system PECR. They lay down the groundwork upon which the formal system PECR can employed in the rigorous construction of computer models of real-world phenomena. In Chapter 10 the computational aspects of proof construction is explored. In the final chapter some derivable rules of PECR along with their proofs are presented.

Notes :: *The formal system PECR has already been outlined in some detail in [9] that was the culmination of some earlier work [27]-[29]. Here we will outline PECR with some important refinements and additional features. We also include a discussion of some recent developments of software where PECR is implemented. It is important to note that the notation used here differs slightly from that used in these earlier works. We have also chosen not to write this article in the stylized typesetting that is typically used in papers of mathematics. We do this so that the alphanumeric character string labels of objects appearing in the text more closely resemble the non-stylized typesetting of actual program code. In an attempt to present this article as a stand alone and complete outline of the subject some material has been directly lifted from [9].*

## 1.3  Current projects

Despite the constraints imposed by the finite resources of a real-world computer it can be shown that abstract theories can be explored on top of our constructive language of computability. To this end one employs programs that construct strings that represent terms. The properties of the terms are conveyed to the machine by way of axioms.

What is of interest here is the recognition that abstractions constructed from a theory based on truth values rely on an underlying validation condition that must ultimately be based on computability. This will not be covered in any detail here since it does not contribute to our main objectives but an interested reader may wish to consult Chapter 9 of [9].

PECR is optimally designed to explore computability of real-world applications and is not the best tool for applications of conventional proof theory. There are many software packages that are available for the latter and a user is better served using these packages for constructing traditional proofs of mathematics.

Projects based on PECR are ongoing and moving towards the ultimate goal of complete automation. PECR has been implemented in the following software packages.

VPC

VPC is a software package that can be employed as an interactive proof assistant as well as a proof checker but not in the conventional sense. VPC is based on CoL as implemented by the formal system PECR. The objective is to explore computability of applications that can be expressed in the language of programs. Particular emphasis is placed on computations that are constrained by the limited resources of the real-world machine on which the application is to be executed.

The original motivation behind the construction of VPC was to explore applications of real-world phenomena by way of computer models employing real-world computation without resorting to the abstractions embedded in conventional theories of computation. While there is an enormous amount of knowledge that has already been acquired in mathematics on the subject of computability, VPC is employed to explore computability from a different perspective. It should be of interest to those who wish to acquire new knowledge under the alternative paradigm of CoL based on real-world computation. Of special interest is the extraction of theories from mathematics that can be translated into a language of programs.

RA

The formal system PECR coupled with machine learning has a stronger association with the scientific method than the formal procedures of mathematics. A preliminary discussion of this topic has been presented in the final chapter of [9]. The software package RA is an attempt to implement these ideas.

The objective here is to construct a fully autonomous machine based rational agent (RA) that explores the laws of nature by observing data of real-world phenomena. RA expresses these laws in a language of programs under the hypothesis that the universe it perceives is computable. RA acquires an understanding of the laws of nature by coupling machine learning with formal reasoning based on CoL as implemented by PECR. At the time of writing of this article RA is in its early stage of development.

Throughout this article we will on occasion refer to VPC and RA in applications. A detailed outline of these software packages along with instructions for setting up applications will be documented elsewhere.

## 1.4  Measure vs patterns

In the current paradigm real-world phenomena are expressed by way of a theory based on the language of mathematics. In this paradigm a computer model is an attempt to approximate the theory with the sole function of being a simulation tool. One of our main objectives is to raise the status of computer models as not only having utility as simulation tools but in themselves are the ultimate expression of the laws that govern real-world phenomena.

We have already discussed the problems faced by describing real-world phenomena in the language of mathematics. If we start with the hypothesis that there exists a real-world that is computable then we can expect that any description of such a real-world must be based on a language of computation that is realizable.

Most contemporary computer models employed in engineering and the sciences are formulated by laws governing certain variables that are assigned values of some measurable quantity. Examples of such variables include mass density, velocity and energy. In the formulations of these models the variables are subject to rules of computation that at the most fundamental level involve the basic operations of arithmetic through addition, subtraction, multiplication and division. In this way we have acquired an understanding of the laws of nature largely from the perspective of variations in quantity.

In recent decades there has been a growing acceptance that our only access to the real-world is through information. Consequently, information becomes fundamental to any model of the real-world. By associating the properties of real-world objects with information the notion of measure becomes less meaningful. Real-world objects can be defined by bits of information that take on the role of labels rather than an expression of quantity. In such models the behavior of real-world phenomena can be viewed as changes in structural patterns that emerge from the distribution of the labels of nodes of a lattice or graph stored as elements of an array.

The information in these arrays are nothing more than labels that can be expressed in the form of alphanumeric character strings or just integers. In this context computer models of real-world phenomena are based on rules of computation that govern the way in which the labels of differing cells of a lattice or elements of an array interact. The rules of arithmetic become less relevant giving way to notions of maps on arrays that focus on the configuration states of the information stored in the arrays.

An example of this approach is found in applications of cellular automata. The simplest applications of cellular automata consist of a lattice where each cell of the lattice is labeled by 0 or 1. The dynamics of a system are governed by rules of cell pair interactions. An extensive exploration of the most elementary rules of cellular automata is covered in [10].

Another example that is, to some extent, related to this area is found in machine learning of images based on neural networks. An image is fed into the neural network in the form of a pattern represented by a matrix of colored cells. To each color is attached a unique numerical value. Combinations of numerical cell values flow through layers of artificial neurons that each computes a weighted sum of the input signals. The process is one of iteration based on an optimization of the weights.

If the image represents a letter or a numeral a rational agent should have no trouble identifying the letter or numeral no matter whether it is highlighted in black on a white background or the color blue on a red background. This suggests that it is not the actual numerical value that quantifies a certain colored cell that matters but rather how the cells are labeled relative to one another. To each color is assigned a unique label. Otherwise the labels can be arbitrary. The primary task of a rational agent is to identify the patterns that emerge from the distributions of the labeled cells in an array.

There is a suggestion here that alternative methods of machine learning could be explored that are based on a more direct approach to pattern recognition that is based on an artificial neuron model that avoids the process of an optimization of weights in a weighted sum. More will be said on this topic elsewhere.

Throughout the course of these notes we will attempt to put more emphasis on computations that focus on the configurations of labels of the elements of arrays from which patterns emerge rather than one based on the basic operations of arithmetic that have traditionally dominated the subject of computer modeling. Nevertheless, our formal system can still be employed to explore the properties of arithmetic. Applications of basic operations of arithmetic on integer scalars and arrays have been covered in earlier works [9] and [27]-[29] using the interactive proof assistant software VPC. Arithmetic of natural numbers on a real-world machine has also been explored by the software package RA and will be documented elsewhere.

## 1.5  Machine parameters

We shall work in a machine environment based on a real-world deterministic computer. A real-world deterministic computer is characterized by the properties of finite information storage along with a finite collection of well defined operations. At any time the machine can exist in any one of a finite number of configuration states. A program is a sequentially ordered list of instructions where each instruction maps the current configuration state of the machine to a new configuration state. The context in which we will choose to work can be defined by the following machine specific parameter constraints.

msym = number of symbols in the alphabet
mstr = maximum number of symbols in any string
mnat = maximum machine number
mlst = maximum number of elements of a list

We define the list of machine parameters by

mach = [msym mstr mnat mlst]

*Throughout we adopt the convention that all lists will be enclosed by square brackets and elements of a list are separated by a space.*

The choice of the parameters of mach are made such that computations are constrained in such a way that they do not make excessive demands on the resources of the machine. The element mlst of the list mach is itself a list that imposes an upper bound on the length of lists of objects of a specific type, e.g. program lists, I/O lists of programs, etc.

We will explore the computational processes that are entirely confined within a machine environment, M, under the specific constraints, mach. Where it is necessary to stress this context we will write

M[mach]

We start by defining the alphabet as a list of symbols

[s[1] s[2] ... s[msym]]

The alphabet that we will work with consists of the following symbols.

Letters:            a b ... z
Digits:             1 2 3 4 5 6 7 8 9 0
Special symbols:    [ ]

A string of the alphabet is a sequence of symbols

s[i[1]]s[i[2]]...s[i[j]]

where j is any number in the range 1 to mstr and the indices i[1],i[2],...,i[j] can be any number in the range 1 to msym. A single string defines an object with generic type denoted by strng.

## 1.6 Types

We shall deal with objects and types, where each object has a type. A machine recognizes each object as a single string or a list of strings.

Object u has type t is denoted by the term

type[u t]

Types may be subtypes of types. Type s is a subtype of type t is denoted by the term

subtype[s t]

Subtypes are transitive, i.e. if subtype[r s] and subtype[s t] then subtype[r t].

A type may also be dependent on other objects. Let b be an object defined by a single string or a list of strings. We write t[b] to mean that the type t depends on the object or list of objects b. Parameter dependent types are subtypes of their generic type, i.e.

subtype[t[b] t]

Note that if s[a] and t[b] are parameter dependent types and subtype[s t] it does not necessarily follow that s[a] is a subtype of t[b].

The main types that will be used in the course of these notes are given in the following table.

*Table 1*

| Alphanumeric label | Description |
|---|---|
| strng | string of symbols (fundamental type) |
| char | alphanumeric string |
| lst | list of strings |
| nat | machine number |
| vec | list of machine numbers |
| arr | array of machine numbers |
| barr | array of elements [0 1] |
| box | bounded regions of type arr objects |
| prgm | program |
| term | term |

An object of type strng is a string made up of any combination of the symbols of the alphabet. An object of type char is an alphanumeric string that is any combination of letters and digits. A machine number is an element of a finite list of natural numbers.

## 1.7 Terms

Terms are strings that have the general form

f[x]

where f is the term name and x is an object of a type that is specific to the term.

Terms will be used to represent the following ::

1. Functions, e.g. f[x] where f is the function name and x is a variable in the form of a scalar, list or array. The variable y=f[x] may be a scalar, list or array.

2. Predicates that test for a relationship between two variables, e.g. r[a b], where r is the term name of the relation between a and b. Relation terms can only be interpreted as being either *defined* in the case that the relation holds or *undefined* if the relation does not hold.

3. Parameter dependent types as described in the previous section.

Functions can be associated with maps. A term, f[x], that represents a function can be associated with a map

map[s t]

where s and t are types. The expression f=map[s t] has the meaning that f maps objects of type s to objects of type t.

A term, f[x], that represents a function is said to be *defined* if its argument x has a value of type s and its target, y=f[x], has a value of type t. The function f[x] is said to be *undefined* for some value assigned x of type s if there is no target value, y=f[x], of type t. A function f[x] is said to be a *partial function* if there is at least one value assignment of x of type s such there exists a target value f[x] of type t. If f[x] can be assigned a value of type t for all value assignments x of type s we say that f[x] is a *total function*.

<u>Notes</u> ::

1. Throughout the course of these notes, terms in combination with natural language will be used to describe functions/maps and relations between objects in an abstract sense and are only meant to represent a preliminary outline of the properties of an object. No attempt will be made in the text to present definitions of objects through the formal style that appears in the mathematical literature. The functional programs that represent terms, i.e. terms are the value assignments of the elements of the I/O lists of a program. The properties of the abstract objects that are expressed as terms can be conveyed to the machine through axioms.

2. In Table 1 the object term is listed as a type. The reason for this is that abstract theories can be explored on top of our formal system based on computation (see Chapter 9 of [9]). For such applications we employ programs that construct strings that represent terms, i.e. terms are the value assignments of the elements of the I/O lists of a program. The properties of the abstract objects that are expressed as terms can be conveyed to the machine through axioms.

## 1.8  Machine numbers

Since we are not interested in computer models that attempt to approximate theories expressed in a language of continuous mathematics we will not make any use of floating point data types. We could accept fixed-point data types because these are essentially integers modified under a scaling factor. However motivated by the ideas discussed in Section 1.4 we will not make explicit use of fixed point data types although they could be included in much of the work presented throughout the course of these notes with some minor modifications of some application specific rules.

When referring to *machine numbers* we will always deal exclusively with objects of type nat. An object of type nat can be assigned any value from the finite list of natural numbers

[0 1 ... mnat]

Object a is type nat is denoted by the term

type[a nat]

To the machine numbers of type nat we associate the usual properties of order. The
following terms for equality and inequality will be used.

*Table 2: Equality and inequality terms :* type[i nat] type[j nat]

| Term | Description |
|------|-------------|
| eqn[i j] | i is equal to j |
| neqn[i j] | i is not equal to j |
| lt[i j] | i is less than j |
| le[i j] | i is less than or equal to j |

We shall sometimes make use of the subtype nat1 of nat.

type[i nat1] states that type[i nat] such that le[1 i]

<u>Notes</u> :: In the earlier works [9] and [27]-[29] machine numbers were defined under
the more general definition of integers that include positive and negative numbers.
There is no loss of generality to our formal system if we restrict the machine
numbers to nonnegative integers.

# 2 Lists

## 2.1 List elements

Objects are represented by a single string or a list of strings. A list of strings is assigned the generic type lst. If u is a list we write

type[u lst]

to mean that u is type lst with unspecified length. The length of a list u is defined as the number of elements in the list and is denoted by

len[u]

The combined statement

type[u lst]
n=len[u]

states that u is a list of length n. This can be stated more concisely as

type[u lst[n]]

where lst[n] is a parameter dependent type meaning type lst of length n. It is understood that lst[n] is a subtype of the generic type lst, i.e.

subtype[lst[n] lst]

The empty list is denoted by

el=[]

An element of a list u at position i, type[i nat1], is denoted by

elt[u i]

Sometimes it will be convenient to express elements in the form of a term that represents a function of its position, i.e.

u[i]=elt[u i]

The list u can be written as a list of its elements

u=[u[1] u[2] ... u[n]]

where each element u[i], i=1,2,...,n, is separated by a space.

If b is an element of the list u we write

in[b u]

## 2.2 Lists of lists

Each element of a list may represent a single string or list of strings. For elements of a list that are themselves lists we use the following notation.

If the element u[i]=elt[u i] of the list u is a list we write

u[i][j]=elt[u[i] j]

to denote the element at position j of the list u[i]. In turn we may have u[i][j]
is also a list, in which case we write a[i][j][k]=elt[a[i][j] k] and so on.

An element of a list that represents a single string that cannot be broken down
into a list of elements will be said to be an *atomic element* of that list.

<u>Example</u> ::

p=[a b], q=[c d e], r=[b a]
u=[p q r]=[[a b] [c d e] [b a]]
v=[r q]=[[b a] [c d e]]
v[1]=elt[v 1]=r=[b a]
v[2]=elt[v 2]=q=[c d e]
v[1][1]=elt[v[1] 1]=b
v[1][2]=elt[v[1] 2]=a
v[2][1]=elt[v[2] 1]=c
v[2][2]=elt[v[2] 2]=d
v[2][3]=elt[v[2] 3]=e

<u>Vectors and arrays</u>

A list whose atomic elements are all integers of type nat will be called a *vector*
and assigned the type vec.

type[u vec[m]]
i=1,2,...,m
    type[elt[u i] nat]

Arrays are multidimensional lists that can be defined by a *dimension list*, d.

type[u lst[d]]
type[d vec[m]]
d=[d[1] d[2] ... d[m]]
i=1,2,...,m
    type[d[i] nat1]

The length, m, of the dimension list, d, is call the *rank* of the array. We write

d=dim[u]

to mean the d is the dimension list of the array u.

A list type lst[m], where m is a scalar and whose elements are all atomic will be
said to be a list of *unit rank*. A vector is a list of unit rank whose atomic
elements are type nat.

*The rank will not be defined for lists whose elements may be lists that are not
structured as arrays.* For example u=[[a b] [a d e]] is a list of lists for which a
dimension list, and hence rank, is not defined.

<u>Index vector</u>

We define the *index vector* by the term, indv[i d], as follows.

```
indv[i d]
type[d vec[m]]
d=[d[1] d[2] ... d[m]]
type[i vec[m]]
i=[i[1] i[2] ... i[m]]
j=1,2,...,m
    le[1 i[j]]
    le[i[j] d[j]]
```

The atomic elements of an array, u, of dimension list, d, can be written as

```
type[u lst[d]]
type[d vec[m]]
indv[i d]
u[i]=u[i[1]][i[2]]...[i[m]]
```

An array whose atomic elements are all integers of type nat is assigned the type arr.

```
type[u arr[d]]
type[d vec[m]]
forall i indv[i d]
    type[u[i] nat]
```

By definition

```
subtype[arr[d] lst[d]]
```

## 2.3  Singleton

By analogy with set theory we import the notion of a *singleton* to mean a list with a single element. In the interests of conserving variable names we will use the same name of the list as the name of the element of the list, i.e. we write u or [u]. The choice of the inclusion of the enclosing brackets will depend on the context, i.e. whether we wish to treat the object as a scalar u or a singleton [u] in the context of lists. In particular, when concatenating a scalar with a list the scalar is treated as a singleton. A singleton u is a list of unit length, i.e. type[u lst[1]].

## 2.4  List operations

<u>List concatenation</u> :: The concatenation of two lists u and v is represented by the term

```
w=conclst[u v]
```

The concatenation, w, of two lists u of length m and v of length n,

```
u=[u[1] u[2] ... u[m]]
v=[v[1] v[2] ... v[n]]
```

has the expanded form

```
type[w lst[m+n]]
w=conclst[u v]=[u[1] u[2] ... u[m] v[1] v[2] ... v[n]]
```

The action of conclst[:] will remove only the outermost brackets that enclose u and v.

<u>Example</u> ::

```
u=[a b c]
v=[d e]
conclst[u v]=[a b c d e]
```

<u>Example</u> ::

```
u=[3 5 2]
v=[2 7]
w=conclst[u v]
 =conclst[[3 5 2] [2 7]]
 =[3 5 2 2 7]
type[w vec[5]]
z=[u v]
 =[[3 5 2] [2 7]]
```

<u>Example</u> ::

List concatenation with scalar a.

```
conclst[a [b c]]=conclst[[a] [b c]]
               =[a b c]
```

<u>Example</u> ::

```
u=[a [b c]]
v=[d e]
z=[u v]=[[a [b c]] [d e]]
conclst[u v]=conclst[[a [b c]] [d e]]
           =[a [b c] d e]
```

The action of conclst[:] removes only the outermost brackets enclosing the lists of its elements. Here the list u has two elements a and [b c].

<u>List intersection</u> :: The intersection of two lists u and v is denoted by the term

```
w=cap[u v]
```

The list w is a list of elements of u that coincide with an element contained in v.

The following algorithm constructs the intersection of two lists.

*Algorithm 1:* `w=cap[u v], type[u lst[n]], type[v lst[m]]`

```
w:=[]
k:=0
do i=1 to n
   if in[u[i] w] cycle
   if in[u[i] v] then
      k:=k+1
      w[k]:=u[i]
   endif
enddo
```

<u>List subtraction</u> :: List subtraction of two lists u and v is denoted by

`w=minus[u v]`

where w contains all elements of u not contained in the list v. By definition

`w=minus[u v]=minus[u cap[u v]]`

The following algorithm constructs w=minus[u v] from type[u lst[n]] and type[v type[m]].

*Algorithm 2:* `w=minus[u v], type[u lst[n]], type[v lst[m]]`

```
w:=[]
k:=0
do i=1 to n
   if in[u[i] v] cycle
   k:=k+1
   w[k]:=u[i]
enddo
```

<u>Repeated elements</u> :: Repeated elements of a list, u, can be removed by the unique function

`w=unique[u]`

where all non repeated elements and only the first element of each repeated element of u is retained in the list w.

The following algorithm constructs w=unique[u] from a list u.

*Algorithm 3:* `w=unique[u], type[u lst[n]]`

```
w:=[]
k:=0
do i=1 to n
   if in[u[i] w] cycle
   k:=k+1
   w[k]:=u[i]
enddo
```

The resultant list, w, of the list operations cap[u v] and minus[u v] maintain the order of the elements as they appear in the first argument of that term, in the above cases the list u. Consequently, the operations cap[u v] and minus[u v] do not satisfy the property of symmetry as a general rule. Note also that for w=cap[u v]
we have w=unique[w]. This is not necessarily the case for w=minus[u v].

<u>Example</u> ::

```
u=[p q r s]
v=[r q s q]
w=[p t]
len[u]=4
len[v]=4
len[w]=2
minus[u w]=[q r s]
a=conclst[u v]=[p q r s r q s q]
unique[a]=[p q r s]=u
cap[u v]=[q r s]
cap[v u]=[r q s]
unique[u]=u
unique[v]=[r q s]
```

<u>Chains</u> :: Repeated use of a concatenation of a list of lists is called a *chain*. If
u, v and w are lists we write

chain[u v w]=conclst[conclst[u v] w]

More generally, if u[1], u[2], ... ,u[n] are lists, not necessarily of equal
length, then

v=chain[u[1] u[2] ... u[n]]

can be constructed by the following algorithm.

*Algorithm 4:* v=chain[u[1] u[2] ... u[n]], type[u[i] lst], i=1,2,...,n

```
v:=u[1]
do i=2 to n
    v:=conclst[v u[i]]
enddo
```

The action of chain[u[1] u[2] ... u[n]] removes only the outermost brackets of each
of its list elements, u[i], i=1,2,...,n.

If each of the lists u[i], i=1,2,...,n, are lists of unit rank, i.e. whose elements
are all atomic, then chain[u[1] u[2] ... u[n]] will be a list of unit rank.

<u>Example</u> ::

```
u=[a b]
v=[b c d]
w=[e f]
y=[u v w]
 =[[a b] [b c d] [e f]]
z=chain[u v w]
 =conclst[conclst[u v] w]
 =conclst[conclst[[a b] [b c d]] [e f]]
 =conclst[[a b b c d] [e f]]
 =[a b b c d e f]
```

<u>Example</u> ::

```
u=[a [b c]]
v=[b d]
w=[e f]
y=[u v w]
 =[[a [b c]] [b d] [e f]]
z=chain[u v w]
 =conclst[conclst[u v] w]
 =conclst[conclst[[a [b c]] [b d]] [e f]]
 =conclst[[a [b c] b d] [e f]]
 =[a [b c] b d e f]
```

Like conclst[:] the action of chain[:] removes only the outermost brackets enclosing the lists of its elements. Here the list u has two elements a and [b c].

<u>List equality</u>

The term

```
eqlst[u v]
```

states that the lists u and v are identical.

The term

```
neqlst[u v]
```

states that the lists u and v are not identical.

## 2.5  Sublists

A list u, type[u lst[m]], type[m nat], is a *sublist* of a list v, type[v lst[n]], type[n nat], and denoted by the term

```
sublst[u v]
```

whenever each element of u is identical to at least one element of v, i.e.

```
sublst[u v]
type[u lst[m]]
type[v lst[n]]
forall i=1 to m
    u[i]=elt[u i]
    in[u[i] v]
```

If sublst[u v] and sublst[v u] then the lists u and v are said to be *equivalent.* This is expressed by the term

```
equivlst[u v]
```

A list u is said to be a *strict sublist* of v if

```
sublst[u v]
neqlst[minus[v u] el]
```

where el is the empty list.

Clearly if eqlst[u v] then equivlst[u v].

<u>Example</u> ::

u=[b b a b]
v=[a b c]
w=[a b a]
sublst[u v]
sublst[w v]
equivlst[u w]

Note that u is a sublist of v even though lt[len[v] len[u]]. Also the two lists u and w are equivalent even though neqlst[u w].

# 3 Programs

## 3.1 Atomic Programs

A program is an object of type prgm. The term

type[p prgm]

states that p is type prgm. Programs are made up of lists of *atomic programs* (APs). APs are assigned the type atm, where

subtype[atm prgm]

The term

type[p atm]

states that p is a program of type atm.

When representing APs we employ labels from the three lists

pname=[pname[1] pname[2] ... pname[nap]]
var=[var[1] var[2] ... var[nvar]]
cst=[cst[1] cst[2] ... cst[ncst]]

The elements of the list pname are the names of the APs. Each element of the input list of an AP is a label taken from elements of var if it is a variable or elements of cst if it is a constant. Labels of elements of the output list of an AP are taken from elements of var only.

An AP, p, is identified by a program name

pn :: program name

I/O lists

x :: list of input elements, type[x lst[nx]]
y :: list of output elements, type[y lst[ny]]

and is represented by the list

p=[pn x y]
 =[pn [x[1] x[2] ... x[nx]] [y[1] y[2] ... y[ny]]]

where x[i]=elt[x i], i=1,2,...,nx, and y[i]=elt[y i], i=1,2,...,ny, are the elements of the I/O lists x and y, respectively. The program name, pn, is a label taken from the list pname.

Each element of an I/O list of a program acts as a placeholder for a *value assignment* (VA). The type of valid VAs of the elements of the I/O lists are specific to the program name, pn.

The I/O lists, x and y, of an AP, p=[pn x y], satisfy the conditions given in Table 3.

*Table 3: I/O list conditions of an atomic program* p=[pn x y]

| Condition 1 | sublst[x conclst[var cst]] sublst[y var] |
|---|---|
| Condition 2 | cap[x y]=[] |
| Condition 3 | y=unique[y] |

Condition 2 states that no element of the input list, x, can have a variable label that coincides with a variable label of the output list, y. Condition 3 states that the variable labels of all elements of the output list are unique.

<u>Terms and programs</u>

Most of the APs that will be presented throughout the course of these notes will be associated with a term. We use the convention that the program name is the same as the associated term name. This can be summarized by the following table.

*Table 4 : Terms and programs*

| Term | Atomic program |
|---|---|
| y=f[x] | p=[f x y] type[p atm] |

A term can be regarded as an abstraction of a program as a function or relation. If the term has the sole task of checking for a type or relation then the output y of the program p is the empty list. For a given VA (value assignment) of the elements of the input list, x, a term is either defined or undefined whereas a program is either computable or uncomputable.

<u>Notes</u> :: The names of programs, elements of I/O lists and types will often be expressed in the form of alphanumeric character strings. There is no special significance in the kind of characters used to represent names of objects. Names of objects will often be referred to as labels. We shall see in Chapter 10 that machine numbers can serve just as well as labels for objects.

## 3.2 Program lists

A program can also be defined as a list of APs. The combined statement

type[p prgm]
n=len[p]

states that p is a program list of APs of length n. These two statements can be combined into the statement

type[p prgm[n]]

where prgm[n] is a parameter dependent type that denotes type prgm of list length n. An AP can also be regarded as type prgm[1]. The APs that make up the program list will sometimes be referred to as *subprograms* of the program list.

A program list is defined by

```
type[p prgm[n]]
p=[p[1] p[2] ... p[n]]
m=1,2,...,n
    p[m]=elt[p m]
    type[p[m] atm]
    p[m]=[pn[m] x[m] y[m]]
```

The I/O lists

```
m=1,2,...,n
    type[x[m] lst[nx[m]]]
    type[y[m] lst[ny[m]]]
```

of each subprogram, p[m], of p have the expanded form

```
x[m]=[x[m][1] x[m][2] ... x[m][nx[m]]]
y[m]=[y[m][1] y[m][2] ... y[m][ny[m]]]
```

Note that x[m] and y[m] now represent lists.

A program list p, type[p prgm[n]], satisfies the conditions listed in Table 5.

*Table 5: I/O dependency conditions of a program list* p,type[p prgm[n]]

| Condition 1 | m=1,2,...,n<br>    elt[p m]=p[m]=[pn[m] x[m] y[m]]<br>    type[p[m] atm]<br>    type[x[m] lst[nx[m]]]<br>    type[y[m] lst[ny[m]]]<br>    in[pn[m] pname]<br>    sublst[x[m] conclst[var cst]]<br>    sublst[y[m] var] |
|---|---|
| Condition 2 | m=1,2,...,n<br>    y[m]=unique[y[m]]<br>    k=1,2,...,n<br>        neqn[k m]<br>        cap[y[m] y[k]]=[] |
| Condition 3 | m=1,2,...,n<br>    k=m,m+1,...,n<br>        cap[x[m] y[k]]=[] |

Conditions 2 and 3 will be referred to as the *I/O dependency conditions*. Condition 2 of Table 5 incorporates Condition 3 of Table 3 and states that all output variable labels of the program list must be unique. Condition 3 of Table 5 states that no element of the input list x[m] of the subprogram p[m] of p can have a variable label that coincides with a variable label of an element of its output list or a variable label of an element of the output list of any subprogram that follows p[m] in the program list p.

A list, p=[p[1] p[2] ... p[n]], that is assigned the value of type prgm will have the expanded representation

```
[[pn[1] x[1] y[1]] [pn[2] x[2] y[2]] ... [pn[n] x[n] y[n]]]
```

We will often express a program as a vertical list. Table 6 summarizes the two equivalent representations of programs.

*Table 6: List representations of a program* `p`, `type[p prgm[n]]`

| Horizontal list | `[[pn[1] x[1] y[1]] [pn[2] x[2] y[2]] ... [pn[n] x[n] y[n]]]` |
|---|---|
| Vertical list | `pn[1] x[1] y[1]`<br>`pn[2] x[2] y[2]`<br>`.          .`<br>`.          .`<br>`.          .`<br>`pn[n] x[n] y[n]` |

When representing a program as a vertical list the outer bounding brackets of each AP will be omitted. The brackets that enclose the I/O lists x and y are retained in both horizontal and vertical lists.

## 3.3 Value assignments

An AP of a program list is represented by a program name followed by its I/O lists. Each element of the I/O list of an AP is a label that acts as a placeholder for a VA (value assignment). If u is the label of an element of an I/O list of a program then we will sometimes write

`val[u]`

to represent the VA of the element u. Each object val[u] may be a single string or a list or array of strings of a type checked within the program. In this way we distinguish the type of the VA of an element of an I/O list from the type of its label taken from the lists var or cst. The VAs of the elements of cst are fixed and need not be of the same type. The list of constants for some applications may be the empty list.

A VA of an element of an I/O list of a program can be thought of as a map that takes each element label of the I/O lists of a program to an object of a specific type.

*Table 7: VA of an I/O list element label,* `u`, *of a program defined as a map*

| `VA:u ↦ val[u]` |
|---|

Each element of the input list is passed into a program as a label taken from the list var or cst. Upon entry all APs check for the type of the VA of each input element. If there are no type violations the AP continues to check a relation or computes a VA of each element of the output list. If the relation or the type of each value assigned output element is consistent with that checked within the program then the program halts and returns the assigned value output, where the output list may be the empty list in the case of a relation checking AP. Otherwise the program halts with an execution error message.

*If the type of the VA of an element of the I/O lists is a subtype of the type allocated for that element in the program then a type violation error will not occur and the internal algorithm of the program proceeds by interpreting the input data as the type defined within the program. Upon execution the VA of the element of the I/O list is returned to its original allocated subtype.*

The list of type names used in an application is given by

```
tname=[tname[1] tname[2] ... tname[ntype]]
```

where each tname[i], i=1,2,...,ntype, is the name of a type.

## 3.4  Variable bindings

The definitions of programs presented in Sections 3.1-3.2 outline the importance of choosing I/O variable labels that avoid violations of the I/O dependency conditions. Repetitions of I/O list element labels are allowed provided they obey certain rules that avoid such conflicts. Elements of I/O lists that have the same label are said to be bound to each other. An element of an input list that has a label taken from the list of constants, cst, is said to be bound to a constant.

The elements of the I/O list, x[m] and y[m], of each AP, p[m]=elt[p m], of the program list, p, type[p prgm[n]], are represented by the notation

```
i=1,2,...,nx[m]
   x[m][i]=elt[x[m] i]

i=1,2,...,ny[m]
   y[m][i]=elt[y[m] i]
```

The allowable element label bindings for the I/O lists are

```
1. le[1 m] le[m n]
   le[1 i] le[i nx[m]]
   in[x[m][i] cst]

2. le[1 m] le[m n]
   le[1 i] le[i nx[m]]
   le[1 k] le[k n]
   le[1 j] le[j nx[k]]
   x[m][i]=x[k][j]

3. le[2 m] le[m n]
   le[1 i] le[i nx[m]]
   le[1 k] le[k m-1]
   le[1 j] le[j ny[k]]
   x[m][i]=y[k][j]
```

It will sometimes be convenient to convert the lists of lists

```
x=[x[1] x[2] ... x[n]]
y=[y[1] y[2] ... y[n]]
```

of a program program p, type[p prgm[n]], into lists of unit rank. We construct the following lists of unit rank

```
inp[p]=chain[x]=chain[x[1] x[2] ... x[n]]
outp[p]=chain[y]=chain[y[1] y[2] ... y[n]]
```

Since each of the elements of x and y are lists of atomic elements (labels) the action of chain[:] converts each of the list of lists x and y into the lists, inp[p] and outp[p], of unit rank (Section 2.4). We can write

```
k=nx[1]+nx[2]+...+nx[n]
l=ny[1]+ny[2]+...+ny[n]
```

```
type[inp[p] lst[k]]
type[outp[p] lst[l]]
```

The action of chain[:] is an unnecessary step if p=[pn x y] is an AP, in which case inp[p]=chain[x]=x and outp[p]=chain[y]=y.

The complete list of distinct I/O element labels that appear in a program, p, is given by

```
lio[p]=unique[conclst[inp[p] outp[p]]
```

where the unique function is used to remove repeated labels.

The intersection

```
cap[outp[p] inp[p]]
```

can be regarded as the list of output variables of a program list, p, that are used as intermediate calculations.

The *primary input list*, pil[p], of a program, p, contains all of those elements of the input lists that are not elements associated with intermediate calculations, i.e.

```
pil[p]=unique[minus[inp[p] outp[p]]]
```

The unique function is used to remove repeated labels.

The list of *free variables*, free[p], of a program, p, contains all elements of the primary input list of p that can be assigned independent values, i.e.

```
free[p]=minus[pil[p] cst]
```

The *primary output list*, pol[p], is defined by

```
pol[p]=minus[outp[p] inp[p]]
```

The list of primary output variables, pol[p], discards output variables that are used as intermediate calculations.

## 3.5  Read/write statements

A program list p, type[p prgm[n]], is executed by first allocating the VAs of its free input variables through a read statement. Constants have fixed VAs that are accessed from memory. Upon execution the VAs of the primary output variables are written to an output file by a write statement. If

```
x'=free[p]
x"=[val[x'[1]] val[x'[2]] … val[x'[len[x']]]
y'=pol[p]
y"=[val[y'[1]] val[y'[2]] … val[y'[len[y']]]
```

then the general format for the execution of a program p, type[p prgm[n]], is given by

```
read [x' f1] [x"]
pn[1] x[1] y[1]
pn[2] x[2] y[2]
  .            .
  .            .
  .            .
pn[n] x[n] y[n]
write [y"] [f2]
```

where f1 is the name of the file where x" is stored and f2 is the name of the destination file where the output y" is to be printed. When constructing a program we will always omit the read and write statements with the understanding that they are present when the program is executed.

The important feature of these read and write statements is that they exclude output variables involved in intermediate calculations and any constants that appear in the input lists of the program, p.

## 3.6  Higher order programs

VAs of elements of I/O lists of programs can be programs. A program whose I/O list elements are assigned values that are type prgm will be called a *higher order program* (HOP). In general, we will say that a HOP is a k-order program if the VAs of its input list elements contain programs of (k-1)-order. A program such that all of its I/O list elements are never assigned values that are type prgm will be called a *zero-order program* (ZOP).

A program that is an element of an I/O list of a HOP is represented by a label taken from the lists var or cst. An element of cst that has a VA of type prgm is said to be a *constant program* and can only appear in the input list of a HOP. If p, type[p prgm], is an element of the I/O lists of a HOP then its VA will take the form of a list of lists as shown in Table 6 of Section 3.2.

Sublists and concatenations of program lists

An element, p, of an I/O list of a HOP represents a list

[p[1] p[2] ... p[n]]

such that the VA of each p[i], i=1,2,...,n, is an AP of the form

val[p[i]]=[pn[i] x[i] y[i]]

A sublist, q, type[q prgm[m]], of the program, p, type[p prgm[n]], has the list representation

q=[q[1] q[2] ... q[m]]

such that for each q[j], le[1 j] le[j m], there exists a p[i], le[1 i] le[i n], such that q[j]=p[i], where equality is based on labels. A program q is a sublist of the program p is denoted by the term

sub[q p]

This is a term specific to program lists and should be distinguished from the generic term sublst[q p] that applies to any lists.

If we have sub[q p] and sub[p q] we say that the two programs are equivalent and write

equiv[q p]

The concatenation of two program lists p, type[p prgm[n]], and q, type[p prgm[m]], is denoted by the term

conc[p q]

and has the expanded representation

[p[1] p[2] ... p[n] q[1] q[2] ... q[m]]

Like program sublists, conc[p q] is a term specific to program concatenations and should be distinguished from the generic term conclst[q p] that applies to any lists.

The programs associated with the terms sub[p q], equiv[p q] and conc[p q] will carry out the same operations as sublst[p q], equivlst[p q] and conclst[p q], respectively, but have the additional task of checking that the elements of their I/O lists have VAs that are type prgm.

## 3.7  Atomic programs of PECR

Tables 8-10 below list the APs of the HOPs that will be used to set up the rules of the formal system, PECR. Some of the terms associated with each program will be defined in subsequent sections.

The name of each AP will also be given an integer label as well as an alphanumeric representation. The first column in the following tables of the APs give the integer labels of program names. The integer labels of program names will be employed in Chapter 10.

*Table 8: Type and relation checking atomic programs*

| Integer label | Alphanumeric representation | Type and relation checks |
|---|---|---|
| 1 | [typep [p] []] | type[p prgm] |
| 2 | [typeap [p] []] | type[p atm] |
| 3 | [equiv [p q] []] | type[p prgm], type[q prgm], equiv[p q] |
| 4 | [sub [q p] []] | type[q prgm], type[p prgm], sub[q p] |
| 5 | [ioeq [q p] []] | type[q prgm], type[p prgm], ioeq[q p] |
| 6 | [ext [p c] []] | type[p prgm], type[c atm], type[c ext[p]] |
| 7 | [flse [p] []] | type[p prgm], type[p false] |

The type checking programs [typep [p] []] and [typeap [p] []] have only one input element and an empty output list and check that the input element, p, has the VA of type prgm and atm, respectively. The relation checking programs named equiv, sub, ioeq and ext have two input elements and an empty output list.

*Table 9: Value and type assignment atomic program*

| Integer label | Alphanumeric representation | Type checks | Value and type assignment |
|---|---|---|---|
| 8 | [conc [p q] [s]] | type[p prgm] type[q prgm] | s:=conc[p q] type[s prgm] |

The program [conc [p q] [s]] first checks that the VAs of p and q are type prgm. It then attempts to construct the program list concatenation s=conc[p q] as defined in the previous section. This may fail if the I/O dependency conditions are violated. If the VA of s=conc[p q] satisfies all of the structural properties of a program it assigns to s the type prgm.

Type assignment programs have the sole task of assigning subtypes of programs. They in effect modify the type of one its input elements, not its assigned value.

*Table 10: Type assignment atomic programs*

| Integer label | Alphanumeric representation | Type checks | Subtype assignment |
|---|---|---|---|
| 9 | [aext [p c] []] | type[p prgm], type[c atm] | type[c ext[p]] |
| 10 | [aflse [p] []] | type[p prgm] | type[p false] |

In the context of HOPs we deal with VAs of I/O list elements of the generic type, prgm, and its subtypes given in Table 11.

*Table 11: Types and subtypes of the VAs of I/O elements of HOPs*

| Alphanumeric label | Object |
|---|---|
| prgm | program |
| atm | atomic program, subtype[atm prgm] |
| ext | program extension, subtype[ext atm] |
| iext | irreducible program extension, subtype[iext ext] |
| false | false program, subtype[false prgm] |
| comp | computable program, subtype[comp prgm] |

All of the objects in the rows below prgm have the generic type prgm. The subtypes ext, iext, false and comp will be defined later. APs are assigned the subtype atm.

<u>Notes</u> :: Programming languages often require that if a list or array is a VA of an element of an I/O list of a functional program then the dimensions of the list or array must also appear in the I/O lists of the program. For example suppose that we wish to concatenate two programs p and q. We could have defined the program [conc [p q] [r]] as [conc [p n q m] [r k]], where n is the list length of the program p and m is the list length of program q. The concatenated program r will have a list length k=n+m.

We can assume that we are working on a machine where the dimensions of a list or array are stored in memory along with the values of their elements. Throughout we adopt a *dimension free format* and assume that the internal algorithm of functional programs can immediately identify the dimensions of a list or array by its name/label along with its element values as they are stored in memory. The omission of parameters associated with list and array dimensions will shorten the lengths of the I/O lists of programs. This is done for brevity only and has very little
bearing on the properties of our formal system.

# 4   Computable Program Extensions

## 4.1  Computability

Terms that represent functions or relations are abstractions of programs. For a given VA of x, a term, f[x], is either defined or undefined. A program replaces this abstraction by either being computable or uncomputable for a given VA of its input list.

Our main objective is to construct programs that can be validated by establishing computability. By this it is meant that a program will eventually halt without encountering an execution error and return a value assigned output, where the output may be the empty list.

Execution error

Within all APs type checking is performed on the assigned values of all elements of its I/O lists. Execution errors are largely based on type violations. A program will halt with an error message if during its execution there is a type violation of any assigned value of the elements of its I/O lists. Otherwise, the execution of a program is completed when all subprograms of a program list have been successfully executed in the sequential order they appear in the program list.

For APs that also check for the satisfaction of a relation between the VAs of a pair of elements of its input list an execution error will also include in the case where the relation is not satisfied.

A program, p, is said to be computable with respect to a VA of its primary input list, pil[p], if upon execution it eventually halts without encountering an execution error. If the program is computable for a given VA of its primary input list, pil[p], it returns a value assigned output, where the output may be the empty list.

A program, p, is assigned the type comp if it is computable for at least one VA of its primary input list. The term

type[p comp]

states that the program p is type comp, where subtype[comp prgm].

Deadline constraint

The above definition of computability requires that a program will 'eventually' halt without encountering an execution error. In some sense this is an abstraction from the perspective of a real-world machine, M[mach]. There are situations where it may be necessary to include in the above definition of computability a deadline constraint. This is particularly the case with RA that couples machine learning with formal reasoning.

With a deadline constraint computability can be tested experimentally by executing a program for a given VA of its primary input list and observing whether it halts with an execution error or returns an assigned value output within the prescribed deadline. In this context programs that do not halt within the deadline are regarded as uncomputable with respect to the VA of its primary input list. While the choice of a deadline constraint is somewhat arbitrary its introduction provides a working framework within which the notion of a *feasible computation* is well defined.

## 4.2  Program extensions

A program, c, is said to be a *program extension* (PE) of a program, p, and is assigned the parameter dependent type ext[p] provided that all of the following conditions are satisfied.

*Table 12: Program extension* type[c ext[p]]

| Condition 1 | s=conc[p c]<br>type[p prgm]<br>type[c atm]<br>type[s prgm] |
|---|---|
| Condition 2 | b=chain[inp[p] outp[p] cst]<br>sublst[inp[c] b] |
| Condition 3 | If p is computable for a VA of its primary input list, pil[p], then the program s=conc[p c] is computable for the same value assigned input |

From Condition 2 it follows that sublst[pil[s] b].

The term ext[p] is a parameter dependent type and

type[c ext[p]]

states that c is type ext[p].

The program s=conc[p c] such that type[c ext[p]] is called an *extended program* (EP) of p.

Notes ::

1. As stated in Condition 1, PEs will always be APs, i.e. for any programs p and c, type[c exp[p]], we have subtype[ext[p] atm].

2. Condition 2 of the definition of a PE states that the input list of c can contain any constant but cannot introduce variable labels that are not contained in the I/O lists of p.

3. Some care is needed not to confuse an EP with a PE. An EP (extended program) of a program p is a program list concatenation s=conc[p c] where c is a PE (program extension) of the program p, i.e. type[c ext[p]].

4. As a generic type ext is least informative as a type since an AP can be a PE of more than one program. Conversely, a program can have more than one extension. The parameter dependent type ext[p] is more informative since it associates the PE to a specific program, p.

Irreducible program extension

A program, c, is an *irreducible program extension* (IPE) of a program, p, and is assigned the parameter dependent type iext[p] provided that all of the following conditions are satisfied.

*Table 13: Irreducible program extension* `type[c iext[p]]`

| Condition 1 | `type[c ext[p]]` |
|---|---|
| Condition 2 | There does not exist a strict sublist q of p such that `type[c ext[q]]` |

The term

`type[c iext[p]]`

states that c is type iext[p]. The program s=conc[p c], `type[c iext[p]]`, is called an *irreducible extended program* (IEP) of p.

## 4.3  Derivation program

The *derivation program* is defined by the program list

```
ext [q c] []
sub [q p] []
conc [p c] [s]
```

Since type[c ext[q]] the input list of c will not introduce variable labels that are not contained in the I/O lists of q. Given in addition q is a sublist of p, it also follows that the input list of c will not introduce variable labels that are not contained in the I/O lists of p. However the I/O dependency conditions (Table 5) also require that the output variable labels of c must not coincide with I/O variable labels of p. The third line of the derivation program has a dual role. It constructs the program list concatenation s=conc[p c] and then checks that s is type prgm.

When applying the program derivation program it will always be possible to choose output variable labels of c that do not violate the I/O dependency conditions.

## 4.4  IEPs and derivations

Our main objective is to construct programs where computability is guaranteed at each step of the program construction. To this end we start with a collection of fundamental rules that will be expressed in the form of IEPs. These fundamental rules will sometimes be referred to as *axioms*. Additional rules, in the form of IEPs, can be derived from these axioms. These will be referred to as *theorems*.

An IEP, s, of a program p, type[p prgm[n]], has the form

```
s=conc[p c]
type[c iext[p]]
```

and can be written as a vertical list

```
pn[1] x[1] y[1]
pn[2] x[2] y[2]
.           .
.           .
.           .
pn[n] x[n] y[n]
---------------
pnc xc yc
```

where the conclusion program, c=[pnc xc yc], type[c atm], is separated from the premise program p, type[p prgm[n]], by a dashed line. Conclusion programs will always be APs.

By the definition of a PE we must have

sublst[xc chain[inp[p] outp[p] cst]]

In order that s=conc[p c] be type prgm we must be also have

yc=unique[yc]

and

cap[yc chain[inp[p] outp[p] cst]]=[]

An IEP can have an empty premise program list, i.e. n=0. In such a case the input list, xc, of c can only contain constants.

Notes ::

1. Here the use of the names *axioms* and *theorems* are borrowed from conventional proof theory. However our notion of an axiom has more of an association with a postulate as it is used in the scientific method. In the sciences a postulate is based on a hypothesis and is assumed to be valid only up until such time that it is found to violate experimental observation. This is an important feature of the software package RA that couples machine learning with formal reasoning.

2. In the text we will, for the most part, separate the premise from the conclusion with a dashed line. However since the conclusion program is always atomic the dashed line is superfluous and has no essential role to play when implementing our formal system on a machine.

## 4.5 I/O equivalence

In Section 3.6 program equivalence was defined in terms of sublists :

**if** sub[p q] **and** sub[q p] **then** equiv[p q]

There is another kind of equivalence that can exist between programs that have different I/O element labels but exhibit some similarity in their I/O element bindings.

Let

type[p prgm[n]]
type[p' prgm[n]]

The program p' is *I/O equivalent* to the program p is expressed by the relation term

ioeq[p' p]

The term ioeq[p' p] is defined if all of the following conditions are satisfied.

*Table 14: I/O equivalence,* `ioeq[p' p]`

| Condition 1 | `n=len[p']=len[p]` |
|---|---|
| Condition 2 | `m=1,2,...,n`<br>   `in[pn[m] pname]`<br>   `in[pn'[m] pname]`<br>   `pn'[m]=pn[m]` |
| Condition 3 | `m=1,2,...,n`<br>`i=1,2,...,nx[m]`<br>   `le[1 k] le[k n]`<br>   `le[1 j] le[j nx[k]]`<br>   **`if`** `x[m][i]=x[k][j]` **`then`** `x'[m][i]=x'[k][j]` |
| Condition 4 | `m=2,3,...,n`<br>`i=1,2,...,nx[m]`<br>   `le[1 k] le[k m-1]`<br>   `le[1 j] le[j ny[k]]`<br>   **`if`** `x[m][i]=y[k][j]` **`then`** `x'[m][i]=y'[k][j]` |
| Condition 5 | `m=1,2,...,n`<br>`i=1,2,...,nx[m]`<br>   **`if`** `in[x[m][i] cst]` **`then`** `x'[m][i]=x[m][i]` |

All equalities in the Table 14 are based on labels. I/O equivalence is reflexive and transitive but not symmetric.

## 4.6  Program extension rule

All axioms are IEPs. The principal axiom of our formal system PECR is called the *program extension rule* and has the axiom label per.

*Program extension rule*

```
Axiom per

ext [q c] []
sub [q p] []
conc [p c] [s]
--------------
aext [p c] []
```

The premise of axiom per is the derivation program. The program extension rule asserts that if c is a PE of the program q and q is a sublist of a program p such that s=conc[p c] is type prgm then c is also a PE of p. The conclusion of axiom per is a type assignment, type[c ext[p]].

## 4.7  Construction rules

The following construction rules of PECR are largely based on program structures. The empty program list is denoted by

ep

and can be regarded as a constant for the formal system PECR.

Once a program has been assigned as a PE it retains that subtype during a program's construction.

*Retention of type program extension*

```
Axiom cr1

aext [p c] []
-------------
ext [p c] []
```

For any PE, type[c ext[p]], there exists an EP, s=conc[p c]. The following rule constructs this EP.

*Construction of an EP*

```
Axiom cr2

ext [p c] []
-------------
conc [p c] [s]
```

*I/O equivalence*

| Axiom cr3a | Axiom cr3b | Axiom cr3c |
|---|---|---|
| typep [p] [] <br> ------------- <br> ioeq [p p] [] | ioeq [p q] [] <br> ioeq [q r] [] <br> ------------- <br> ioeq [p r] [] | ext [q c] [] <br> conc [q c] [r] <br> conc [p d] [s] <br> ioeq [p q] [] <br> ioeq [s r] [] <br> ------------- <br> aext [p d] [] |

*Program sublists*

| Axiom cr4a | Axiom cr4b | Axiom cr4c |
|---|---|---|
| typep [p] [] <br> ------------- <br> sub [p p] [] | sub [p q] [] <br> sub [q p] [] <br> ------------- <br> equiv [p q] [] | sub [p q] [] <br> sub [q r] [] <br> ------------- <br> sub [p r] [] |

*Sublists of program list concatenations*

| Axiom cr5a | Axiom cr5b | Axiom cr5c |
|---|---|---|
| conc [p q] [s] <br> ------------- <br> sub [p s] [] | conc [p q] [s] <br> ------------- <br> sub [q s] [] | conc [p q] [s] <br> sub [p r] [] <br> sub [q r] [] <br> ------------- <br> sub [s r] [] |

*Program list concatenation with the empty program*

| Axiom cr6a | Axiom cr6b | Axiom cr6c |
|---|---|---|
| typep [p] [] <br> --------------- <br> conc [p ep] [s] | typep [p] [] <br> --------------- <br> conc [ep p] [s] | conc [p ep] [s] <br> --------------- <br> equiv [s p] [] |

*Program subtype*

| Axiom cr7 |
|---|
| typeap [a] [] <br> ------------- <br> typep [a] [] |

The principal axiom per, the construction rules, cr1-cr7, along with the axioms of falsity, flse1-flse2, the IOT axiom and the substitution rule form the main core of our formal system PECR (Program Extension Construction Rules). Formal definitions of axioms flse1-flse2, the IOT axiom and the substitution rule will be provided later.

Notes ::

1. In axiom cr6c we have extended the definition of program equivalence to include program list concatenations with the empty program.

2. The substitution rule under equivalence can be applied as an axiom to the programs named sub, equiv and conc.

## 4.8  Derivable rules

We deal with objects that are subtypes of strings or lists of strings such as scalars, vectors, arrays of machine numbers and programs. All of these objects have a well defined string structure as specified by their definitions and are recognized by the machine.

The program [ext [p c] []] checks that p and c are type prgm and c is type ext[p]. A machine can readily verify Conditions 1 and 2 of Table 12 from the string structures of p and c. However from the perspective of a feasible computation, the machine has no general way of recognizing that a concatenation of two program lists has the property associated with computability as outlined in Condition 3 of Table 12. Consequently a machine can only interpret objects of type ext through the properties embedded in the construction rules. In particular a program c can only acquire the subtype ext[p], for some program p, through the type assignment program [aext [p c] []] or be defined as such through axioms.

The substitution rule under equivalence should not be applied to the programs [ext [p c] []] and [aext [p c] []] as an axiom. However it will be shown in Chapter 11 that with the inclusion of the supplemental axiom eope (Section 6.2) the AP [ext [p c] []] does satisfy the substitution rule under equivalence as a derivation.

As stated in axioms cr3a-cr3b, I/O equivalence is reflexive and transitive. I/O equivalence is not symmetric. The substitution rule under program equivalence should not be applied as an axiom to the I/O equivalence checking program [ioeq [p q] []].

There are several rules that appear to be fundamental but are derivable from the axioms of the previous section. For instance program equivalence is reflexive, symmetric and transitive, i.e.

| type [p] [] | equiv [p q] [] | equiv [p q] [] |
|---|---|---|
| -------------- | -------------- | equiv [q r] [] |
| equiv [p p] [] | equiv [q p] [] | -------------- |
| | | equiv [p r] [] |

These are not axioms. The first two rules can be derived and the third rule follows immediately from the substitution rule under equivalence. Derivations of these and other rules are presented in Chapter 11.

## 4.9  Natural deduction

The statement type[c ext[p]] asserts that whenever the program p is computable for a VA of its primary input list, pil[p], the AP c is computable for the same VA of the EP conc[p c]. In proof theory, judgments of natural deduction take the form

$$\Pi \vdash C$$

where $C$ is a single formula and

$$\Pi = P_1, P_2, \ldots, P_n$$

is a sequent of formulas, $P_i$, i=1,2,...,n. Here the symbol $\vdash$ is used to represent entailment and we say that $\Pi$ entails $C$. The semantics of the expression $\Pi \vdash C$ asserts that whenever all of the formulas $P_1, P_2, \ldots, P_n$ are true then $C$ is true.

In our formal system, PECR, we use instead of the sequent $\Pi = P_1, P_2, \ldots, P_n$ the program list p=[p[1] p[2] ... p[n]] and instead of the formula $C$ the atomic program c. The statement of entailment, $\Pi \vdash C$, is then replaced by the statement type[c ext[p]]. The following table summarizes the analogy between natural deduction and program extensions.

*Table 15: Analogy between Natural Deduction and PEs in PECR.*

| | **Natural Deduction** | **PECR** |
|---|---|---|
| **Premise** | *Sequent* :: $\Pi = P_1, P_2, \ldots, P_n$ | *Program list* :: p=[p[1] p[2] ... p[n]] |
| **Conclusion** | *Formula* :: $C$ | *Atomic program* :: c |
| | *Entailment* :: $\Pi \vdash C$ | *Program extension* :: type[c ext[p]] |

## 4.10    False programs

There are cases where an object has all of the structural properties of a program as outlined in Sections 3.1-3.2 but will be uncomputable for any type compatible VA of its primary input list.

A program is said to be a *false program* if there does not exist a VA of its primary input list such that the program is computable. The term

type[p false]

states that the program p is a false program.

By definition we have

subtype[false prgm]

This means that for an object to have the type assignment false it must first have the structure of a program as outlined in Sections 3.1-3.2. A program that will always halt as a result of an error in syntax is not considered meaningful in this context since such an object cannot be assigned the type prgm.

The HOP

[flse [p] []]

checks that the program p has been assigned the type false. A program will automatically be assigned the type false if it is defined as such as an axiom of falsity. Otherwise a program, p, can only be assigned the type false through the type assignment program

[aflse [p] []]

A program p, type[p false], is said to be an *irreducible false program* if there does not exist a program q such that q is a strict sublist of p such that type[q false].

To the construction rules presented in the Section 4.7 we introduce the additional rules of falsity.

*Sublist falsity rule*

```
Axiom flse1

sub [q p] []
flse [q] []
------------
aflse [p] []
```

*Retention of subtype assignment*

```
Axiom flse2

aflse [p] []
------------
flse [p] []
```

<u>Axioms of falsity</u>

In applications an IEP of falsity for a program p, type[p prgm[n]] and type[p false], can be stated as

```
pn[1] x[1] y[1]
pn[2] x[2] y[2]
 .          .
 .          .
 .          .
pn[n] x[n] y[n]
---------------
false
```

In the context of HOPs the AP [flse [p] []] is a type checking program that checks that the program p has been assigned the type of a false program. An axiom of falsity can also be regarded as a higher order construct of an IEP with an empty list premise and can be written as

```
-----------
flse [p] []
```

*The lower-order program,* p, *that appears as input to the conclusion program of the axiom of falsity, is application specific. As such it must be defined as a constant of a type* prgm *object in the context of a HOP.*

Note that while a program, p, type[p false], is not computable for any VA of its primary input list, the HOP [flse [p] []] is computable.

<u>Notes</u> :: The substitution rule under equivalence should not be applied as an axiom to the programs [flse [p] []] and [aflse [p] []]. However, as demonstrated in Chapter 11, the program [flse [p] []] does satisfy the substitution rule under equivalence as a derivation.

## 4.11      Connection List

Let p=[p[1] p[2] ... p[n]] be a proof program with the first m, lt[m n], statements [p[1] p[2] ... p[m]] being the premise of the proof. When a proof is completed VPC and RA will output the proof program as a vertical list with the first m lines written as

```
i p[i]
```

where i=1,2,,...,m, is the statement label (line number) and the AP p[i] is a statement of the premise.

Each of the lines i=m+1, m+2,...,n, are written as

```
i p[i] atl clist
```

The derived statement p[i] is followed by the IEP label, atl, that was employed in the derivation of the statement p[i] and clist is the *connection list*.

If k is the length of the premise of the IEP labeled atl then the connection list is of the form

```
clist=[l[1] l[2] ... l[k]]
```

where le[l[j] i-1], j=1,2,...,k, are the line numbers of the sublist of p,

`[p[l[1]] p[l[2]] ... p[l[k]]]`

that is I/O equivalent to premise of the IEP labeled atl that was employed in the derivation of the statement p[i].

# 5   IOT and substitution rules

Along with the program extension rule, per, the construction rules cr1-cr7 and the axioms of falsity fle1-flse2 there are two more rules that form the main core of the formal system PECR. These additional rules are the IOT axiom and the substitution rule. These are automated in VPC and RA.

## 5.1  IOT axiom

All APs internally check for type violations of the VAs of the elements of their I/O lists. APs whose sole task is to check the type of the VA of an I/O element are generally given the name

type*X*

where *X* represents an alphanumeric substring that is associated with the type being checked. The input list has only one element and the output list is empty. The type checking program halts with an execution error message if there is a type violation.

In this article the type checking programs used are given in the following table.

| Type checking program | Term | Description |
|---|---|---|
| [typen [a] []] | type[a nat] | type machine number |
| [typea [a] []] | type[a arr] | type array of machine numbers |
| [typebx [a] []] | type[a box] | type box |
| [typep [a] []] | type[a prgm] | type program |
| [typeap [a] []] | type[a atm] | type atomic program |

An AP, p, has the list representation [pn x y]. The elements of the I/O lists are given by x[i]=elt[x i], i=1,2,...,nx, and y[j]=elt[y j], j=1,2,...,ny.

The IOT axiom is applied to both input and output list elements and can be expressed as follows.

```
Axiom iot

i=1,2,...,nx

pn x y
---------------
typeX [x[i]] []

j=1,2,...,ny

pn x y
---------------
typeY [y[j]] []
```

Here the substrings *X* and *Y* depend on the type of the VA of each I/O element associated with the AP [pn x y]. The IOT axiom states that if a program is computable for a given VA of its I/O elements then the VA for each I/O element will check positively under the appropriate type checking program. Both VPC and RA will identify the appropriate type checking programs, type*X* and type*Y*, based on the type of the VA of each x[i] and y[j], respectively, as allocated in the AP [pn x y].

## 5.2  Substitution rule

Excluding I/O equivalence, equality/equivalence program names have the general form

eq*X*

where *X* represents an alphanumeric character substring that corresponds to the type of the VAs of the input elements. Equality/equivalence programs have two input elements and like type checking programs have an empty output list.

Equality/equivalence programs first check that the VAs of their two input elements have the same type that are specific to the equality program name and then checks that they are equal/equivalent. The program halts with an execution error message if there is a type or an equality/equivalence violation.

Substitution rule under equality

In this article the main equality checking programs used are given by the following table.

| Equality checking programs | Term | Description |
|---|---|---|
| [eqn [a b] []] | eqn[a b] | Equality of machine numbers |
| [eqa [a b] []] | eqa[a b] | Equality of arrays of machine numbers |
| [eqbx [a b] []] | eqbx[a b] | Equality of boxes |

Let p and p' be two APs with the list representations [pn x y] and [pn x' y'], respectively, such that with respect to variable labeling

cap[y y']=[]

and for some k

i=1,2,...,nx
    neqn[k i]
    x'[i]=x[i]

i.e. the elements of the input lists x and x' have the same labels except possibly for the elements x[k] and x'[k]. In other words the lists minus[x [x[k]]] and minus[x' [x'[k]]] are identical.

The substitution rule comes in two parts. The first part is an existence axiom.

*Substitution rule 1*

| Axiom sr1 |
|---|
| |
| pn x y |
| eq*X* [x'[k] x[k]] [] |
| -------------------- |
| pn x' y' |

Clearly the substitution rule is satisfied when the elements x[k] and x'[k] have the same labels.

The second part of the substitution rule is an equality axiom and is applicable to programs with nonempty output lists.

*Substitution rule 2*

```
┌─────────────────────────────────┐
│ Axiom sr2                       │
├─────────────────────────────────┤
│                                 │
│ j=1,2,...,ny                    │
│                                 │
│ pn x y                          │
│ eqX [x'[k] x[k]] []             │
│ pn x' y'                        │
│ --------------------            │
│ eqY [y'[j] y[j]] []             │
└─────────────────────────────────┘
```

where the second part, sr2, is generated for each element pair y'[j] and y[j], j=1,2,...,ny, of the output lists. Here we use the equality names eqX and eqY, where *X* and *Y* may be the same or different substrings depending on the type of the VAs of the elements of the I/O lists. Both VPC and RA will identify the appropriate program, eqY, based on the type of the VA of each y[j].

<u>Example</u> :: In Chapter 9 we introduce the program [box [a b] [p]] that constructs the multidimensional discrete box p=[a b], type[p box]. The VAs of a and b are arrays, type arr, that represent the bounds of the box p. The substitution rule applied to the first input element of the program [box [a b] [p]] under equality is given by

```
box [a b] [p]                    box [a b] [p]
eqa [c a] []                     eqa [c a] []
-------------                    box [c b] [q]
box [c b] [q]                    -------------
                                 eqbx [q p] []
```

Note that in second part of the substitution rule there are two equality programs [eqa [a c] []] and [eqbx [q p] []], the first associated with equality of arrays and the second associated with equality of boxes.

<u>Substitution rule under equivalence</u>

The substitution rule can be applied to substitutions under equivalence. However substitution under equality and equivalence cannot be mixed in axiom sr2, i.e. the two occurrences of the program names eqX and eqY must either be both equality program names or both equivalence program names.

<u>Example</u> ::

```
conc [p q] [s]                   conc [p q] [s]
equiv [r q] []                   equiv [r q] []
--------------                   conc [p r] [t]
conc [p r] [t]                   --------------
                                 equiv [t s] []
```

<u>Example</u> ::

```
sub [q p] []
equiv [r q] []
--------------
sub [r p] []
```

Only the first part of the substitution rule, sr1, is applicable here since [sub [q p] []] has an empty output list.

<u>Notes</u> :: Some APs can be shown to satisfy the substitution rule under equality/equivalence from the other axioms. The substitution rule should not be regarded as an axiom for such programs. When setting up an application in VPC a user is required to supply the names of the programs for which the substitution rule under equality/equivalence is to be applied as an axiom.

## 5.3  Identity rule

Repetitions of subprograms in a program list are allowed for subprograms with an empty output list. Sometimes we may wish to have a repetition of a program with a nonempty output list, but this is only possible by introducing new labels for the output list variables.

The following rules will hold in general.

```
pn x y                              j=1,2,...,ny
-------
pn x y'                             pn x y
                                    pn x y'
                                    -------------------
                                    eqY [y'[j] y[j]] []
```

where y' is a relabeling of the elements of y. These rules are not axioms because they can be derived by employing the program [eqX [x[k] x[k]] []] in the substitution rules sr1 and sr2.

These two rules combined can be interpreted as an identity rule. An example where repetitions of programs with nonempty output lists are employed in proofs can be found in theorem thm1 presented in Section 9.6.

# 6 Disjunctions and conjunctions

## 6.1 Subtypes of atomic programs

The program extension rule, per, along with the construction rules cr1-cr7, flse1-flse2, the IOT axiom and the substitution rule form the main core of the formal system PECR. In this chapter we will introduce disjunctions and conjunctions that can sometimes be useful in certain applications. The rules associated with disjunctions and conjunctions can be regarded as supplemental to the main core axioms of PECR.

A *disjunction*, d, of two programs a and b will be represented by the term

d=disj[a b]

where the programs a and b are called the operands of the disjunction program d.

A *conjunction*, c, of two APs a and b will be represented by the term

c=conj[a b]

where the internal algorithm of the conjunction program c is a program list concatenation of the two APs a and b.

Disjunctions and conjunctions will be regarded as APs and can be employed as PEs.

All PEs must be of type atm and by definition are single list programs, i.e. type prgm[1]. There are three subtypes of atm.

*Table 16: Subtypes of* atm

| AP subtype | Description |
|---|---|
| fatm | *Fundamental atomic program :* The internal code of an AP of subtype fatm is written in some imperative language. Each AP of subtype fatm may call other APs but must introduce a unique computation. |
| dsj | *Disjunction :* Constructed from two programs, each referred to as an operand of the disjunction. |
| cnj | *Conjunction :* Constructed from program list concatenations of two APs. |

To the APs of Tables 8-10 we include the disjunction and conjunction APs.

*Table 17: Value and type assignment atomic programs of disjunctions and conjunctions*

| Integer label | Alphanumeric representation | Type and relation checks | Value and type assignment |
|---|---|---|---|
| 11 | [disj [p q] [s]] | type[p prgm], type[q prgm] equivlst[free[p] free[q]] equivlst[pol[p] pol[q]] | s:=disj[p q] type[s atm] |
| 12 | [conj [p q] [s]] | type[p atm], type[q atm] type[conc[p q] prgm] | s:=conj[p q] type[s atm] |

The elements of the input list of conjunctions have VAs that are type atm as distinct from the generic type prgm. The output element of both disjunctions and conjunctions have a VA of type atm.

<u>Notes</u> :: We have used the names disjunctions and conjunctions for the program constructions [disj [a b] [d]] and [conj [a b] [c]] because they have some similarities with disjunctions and conjunctions as they are defined in classical logic. However these program constructions will possess certain properties that are not reflected in classical logic.

## 6.2  Program equivalence

So far program equivalence has been defined in terms of program sublists and program list concatenations with the empty program. Roughly speaking the notion of program equivalence is motivated by the property that nonidentical programs produce the same value assigned output for the same value assigned input. A more precise definition of program equivalence can be stated as follows.

The programs p and q will be said to be program equivalent if

equivlst[free[p] free[q]]
equivlst[pol[p] pol[q]]

and

if

**forall** i=1 **to** len[free[p]]
   **exists** free[q][k]=free[p][i]
   val[free[q][k]]=val[free[p][i]]

then

**forall** j=1 **to** len[pol[p]]
   **exists** pol[q][l]=pol[p][j]
   val[pol[q][l]]=val[pol[p][j]]

The equalities in the above existence statements are associated with I/O element labels while the equalities following the existence statements are associated with VAs. Using this definition we will extend the property of program equivalence to include disjunctions and conjunctions.

Two programs u and v are program equivalent and denoted by the term

equiv[u v]

if any of the following conditions are met:

   1. sub[u v]
      sub[v u]

   2. u=conc[v ep]

   3. u=disj[a b]
      v=disj[b a]

4. u=disj[v b]
   type[b false]

5. d=disj[a b]
   u=conc[p d]
   v=disj[conc[p a] conc[p b]]

6. d=disj[a b]
   u=conc[d p]
   v=disj[conc[a p] conc[b p]]

7. u=conc[conc[typeap[a] typeap[b]] conc[a b]]
   v=conj[a b]

In our formal system a PE is always an AP. Under our extended definition of program equivalence the following axiom will hold for program equivalence of PEs that include all of the AP subtypes fatm, dsj and cnj.

*Equivalence of program extensions*

```
Axiom eope

ext [p c] []
equiv [b c] []
--------------
aext [p b] []
```

## 6.3  Disjunctions

A disjunction, p, is a program represented by the term

p=disj[pa pb]

where pa and pb are programs that are referred to as the *operands* of the disjunction p. A disjunction has type dsj, where subtype[dsj atm].

For a given VA of its input list a disjunction is computable if at least one of its operand programs is computable. Otherwise the disjunction is uncomputable.

Given

type[pa prgm]
type[pb prgm]

such that

equivlst[free[pa] free[pb]]
equivlst[pol[pa] pol[pb]]

the HOP

[disj [pa pb] [p]]

constructs the program disjunction p, type[p dsj], such that

```
val[pa]=[pna xa ya]
val[pb]=[pnb xb yb]
val[p]=[pn x y]

x=free[pa]
y=pol[pa]
```

We are using the convention that the I/O lists of a disjunction, p=disj[pa pb], are determined by the free variables and primary output list of the first operand program pa. This means that the disjunction disj[pb pa] need not be identical to p. However both disjunctions are program equivalent as stated by axiom dsj2b below.

## Disjunction axioms

The first disjunction axiom states that the coinciding PEs of the operands can be assigned as a PE of the disjunction.

*Disjunction contraction rule 1*

```
┌─────────────────────────────┐
│ Axiom dsj1                  │
├─────────────────────────────┤
│                             │
│ ext [a c] []                │
│ ext [b c] []                │
│ disj [a b] [d]              │
│ --------------              │
│ aext [d c] []               │
└─────────────────────────────┘
```

The following disjunction contraction rules involve false programs. They are not stated as axioms because they can be derived (Chapter 11).

*Disjunction contraction rule 2*

```
flse [a] []
ext [b c] []
disj [a b] [d]
--------------
aext [d c] []
```

*Disjunction contraction rule 3*

```
flse [a] []
flse [b] []
disj [a b] [d]
--------------
aflse [d] []
```

*Disjunction commutativity*

| Axiom dsj2a | Axiom dsj2b |
|---|---|
| disj [a b] [s]<br>--------------<br>disj [b a] [r] | disj [a b] [s]<br>disj [b a] [r]<br>--------------<br>equiv [r s] [] |

*Disjunction distributivity (right)*

| Axiom dsj3a | Axiom dsj3b |
|---|---|
| disj [a b] [d]<br>conc [p a] [r]<br>conc [p b] [s]<br>--------------<br>disj [r s] [v] | disj [a b] [d]<br>conc [p a] [r]<br>conc [p b] [s]<br>conc [p d] [u]<br>disj [r s] [v]<br>--------------<br>equiv [u v] [] |

*Disjunction distributivity (left)*

| Axiom dsj4a | Axiom dsj4b |
|---|---|
| disj [a b] [d]<br>conc [a p] [r]<br>conc [b p] [s]<br>--------------<br>disj [r s] [v] | disj [a b] [d]<br>conc [a p] [r]<br>conc [b p] [s]<br>conc [d p] [u]<br>disj [r s] [v]<br>--------------<br>equiv [u v] [] |

*False operand program*

| Axiom dsj5 |
|---|
| disj [a b] [d]<br>flse [b] []<br>--------------<br>equiv [d a] [] |

*Extension disjunction introduction*

| Axiom dsj6 |
|---|
| ext [p a] []<br>disj [a b] [c]<br>conc [p c] [s]<br>--------------<br>aext [p c] [] |

We can assume that disjunctions satisfy the substitution rule under program equivalence.

## 6.4 Conjunctions

The internal algorithms of APs of subtype fatm can be written in some imperative language. An AP of subtype fatm can call other APs but must introduce a new computation. A conjunction is a single list program whose internal algorithm is the program list concatenation of two APs. Conjunctions are treated as APs in
derivations and are allocated the type cnj, where subtype[cnj atm].

A conjunction, p, is represented by the term

p=conj[pa pb]

whose internal algorithm is the program list concatenation of the two APs pa and
pb. It is distinguished from the standard program list concatenation as follows.

Given

type[pa atm]
type[pb atm]
[conc[pa pb] [s]]
type[s prgm]

the HOP

[conj [pa pb] [p]]

constructs the program conjunction p, type[p cnj], such that

val[pa]=[pna xa ya]
val[pb]=[pnb xb yb]
val[p]=[pn x y]

x=free[s]
y=pol[s]

In this way the I/O lists of p remove all constants and output variables used as
intermediate calculations in the I/O lists of the program concatenation s.

<u>Conjunction axioms</u>

| Axiom cnj1 |
|---|
| ext [p a] []<br>ext [p b] []<br>conj [a b] [c]<br>--------------<br>aext [p c] [] |

| Axiom cnj2a | Axiom cnj2b | Axiom cnj2c |
|---|---|---|
| conj [a b] [c]<br>--------------<br>conc [a b] [d] | typeap [a] []<br>typeap [b] []<br>conc [a b] [c]<br>--------------<br>conj [a b] [d] | conj [a b] [c]<br>conc [a b] [d]<br>--------------<br>equiv [d c] [] |

Axiom cnj2a reflects the property that [conj [a b] [c]] internally checks that the
program list concatenation conc[a b] is type prgm.

The presence of the conjunction program [conj [a b] [c]] in the premises of cnj2a and cnj2c means that the inclusion of the type checking programs [typeap [a] []] and [typeap [b] []] are superfluous because they are internally called by [conj [a b] [c]]. This is reflected in the application of the IOT axiom to the conjunction program [conj [a b] [c]].

```
conj [a b] [c]            conj [a b] [c]            conj [a b] [c]
--------------            --------------            --------------
typeap [a] []             typeap [b] []             typeap [c] []
```

APs of subtype cnj can be built up from the recursive binary operations of program list concatenations of APs of subtype fatm, dsj and cnj. Conjunctions need to be accompanied by application specific axioms (Section 7.4).

We can assume that conjunctions satisfy the substitution rule under program equivalence.

<u>Notes</u> :: APs of subtype fatm are fundamental in the sense that each introduces a new computation. Conjunctions are internally constructed from the program list concatenations of pairs of APs so it is more appropriate to regard them as pseudo-atomic programs.

# 7   Applications

## 7.1  Applications of PECR

An *application* in PECR replaces the traditional notion of a *theory* as it is currently used in the sciences and mathematics. An application comes with a list of its own APs whose names are given by the list

pname=[pname[1] pname[2] ... pname[nap]]

a list of axioms

ax=[ax[1] ax[2] ... ax[nax]]

and a list of constants

cst=[cst[1] cst[2] ... cst[ncst]]

An application is also accompanied by a list of types.

tname=[tname[1] tname[2] ... tname[ntype]]

Each element, ax[i]=elt[ax i], i=1,2,...,nax, of the list, ax, is an IEP of the form conc[p c], type[c iext[p]]. We write

*S*[pname ax mach]

to represent the application *S* with its dependencies that include the machine parameters, mach (Section 1.5). Here *S* is an alphanumeric character string that represents the name of the application.

The list of all IEPs that exist in an application *S*[pname ax mach] can be partitioned into the list s given by

s=[ax th ud]

where ax is the list of axioms of the application *S*, th is the list of IEPs that have proofs that can be derived from the axioms of *S* (theorems), and ud is the list of all IEPs that have no proofs that can be derived from the axioms of *S*.

Elements of ax and ud are IEPs that cannot be derived. The elements of ax are distinguished from elements of ud by the property that for each element of ax there exists at least one element of th whose proof contains a dependence on that axiom. A desirable feature of any application, *S*, is that the size of the lists ax and ud are minimal. An application, *S*, is said to be *complete* if the list ud is the empty list. An application, *S*, such that ud is a nonempty list is said to be *incomplete*.

Notes :: Each element of the lists ax, th and ud is type prgm but the lists themselves are not meant to be type prgm.

## 7.2  Special applications

An application is denoted by the term *S*[pname ax mach] where the lists pname, ax, and mach are specific to that application. In this article we will sometimes refer to the applications denoted by the names *S*=pecr and *S*=nat.

1. nat[pname ax mach] :: This application can be divided into three main parts. In the first part we will present some basic order axioms for type nat objects. In the second part we will explore discrete intervals and their extension to multidimensional boxes. Having examined some basic properties of boxes we then proceed to the exploration of computability of *fully discrete dynamical systems*. The properties of fully discrete dynamical systems lay down the foundations from which computer models of real-world phenomena can be rigorously constructed. The APs and axioms of arrays, boxes and dynamical systems are presented in Chapter 9.

2. pecr[pname ax mach] :: The formal system PECR can be regarded as an application of itself. We start with the axioms of PECR and derive additional HOP rules (Chapter 11). In this application pname is the list of AP names given in Tables 8–10 and Table 17 and ax is the list containing the program extension rule, per, the construction rules cr1-cr7, the axioms of falsity flse1-flse2 along with the supplemental axioms for disjunctions and conjunctions, eope, dsj1-dsj6 and cnj1-cnj2.

For demonstration purposes proofs of a few theorems of the applications nat[pname ax mach] and pecr[pname ax mach] will be presented. VPC has been employed to explore several other applications that include the basic operations of vectors and arrays as well as abstract theories that address completeness. The theorems and their proofs can be found in [9] and [27]-[29].

## 7.3  Application specific axioms of disjunctions

Given

type[pa prgm]
type[pb prgm]

such that

equivlst[free[pa] free[pb]]
equivlst[pol[pa] pol[pb]]

the HOP

[disj [pa pb] [p]]

constructs the program disjunction p, type[p dsj], such that

val[pa]=[pna xa ya]
val[pb]=pnb xb yb]
val[p]=[pn x y]

x=free[pa]
y=pol[pa]

The program name, pn, of the disjunction p must be included in the application specific list of AP names, pname.

Let

na=len[pa]
nb=len[pb]

```
len[y]=len[y']
cap[y y']=[]
```

where y' is a relabeling of the elements of y.

The application specific disjunction axioms take the form

```
pna[1] xa[1] ya[1]                    pnb[1] xb[1] yb[1]
pna[2] xa[2] ya[2]                    pnb[2] xb[2] yb[2]
.                                     .
.                                     .
.                                     .
pna[na] xa[na] ya[na]                 pnb[nb] xb[nb] yb[nb]
---------------------                 ---------------------
pn x y'                               pn x y'
```

If y is a nonempty list then equality axioms can be included that link VAs of the elements of y' with the associated output list elements of the operand programs.

The application specific axioms of disjunctions are not automated in VPC and must be user supplied and included along with the other axioms that are contained in an initializing file. A cautionary note here is that some of these rules may be derivable from the other axioms of the application. Such rules should be omitted from the initializing file and derived when needed. An example of this can be found in the trichotomy program presented in Section 8.4.

## 7.4  Application specific axioms of conjunctions

Given

```
type[pa atm]
type[pb atm]
[conc[pa pb] [s]]
type[s prgm]
```

the HOP

```
[conj [pa pb] [p]]
```

constructs the program conjunction p, type[p cnj], such that

```
val[pa]=[pna xa ya]
val[pb]=[pnb xb yb]
val[p]=[pn x y]
```

```
x=free[s]
y=pol[s]
```

The program name, pn, of the conjunction p must be included in the application specific list of AP names, pname.

There are two kinds of application specific axioms for conjunctions. The first kind state that if both of the programs [pna xa ya] and [pnb xb yb] are computable then so is the conjunction [pn x y].

The second kind state that if a conjunction [pn x y] is computable then each of the programs [pna xa ya] and [pnb xb yb] are computable. Some care is required when constructing application specific axioms of the second kind. If cap[xb ya] is not the empty list then one must include the extra conditional statement [pna xa ya] in the premise to establish the computability of [pnb xb yb] (see third example below).

Like disjunctions, application specific axioms of conjunctions are not automated in VPC and must be user supplied and included along with the other axioms that are contained in an initializing file.

The following examples construct conjunctions from the APs of Chapter 9.

<u>Example</u> :: We construct the conjunction [conj [pa pb] [p]] where

val[pa]=[lea [a v] []]
val[pb]=[lea [v b] []]
val[p]=[pn [a v b] []]

The application specific axioms of the first kind for this conjunction state that if the APs [lea [a v] []] and [lea [v b] []] are computable then so is the conjunction [pn [a v b] []], i.e.

lea [a v] []
lea [v b] []
-------------
pn [a v b] []

The application specific axioms of the second kind for this conjunction state that if the conjunction [pn [a v b] []] is computable then so are the APs [lea [a v] []] and [lea [v b] []], i.e.

pn [a v b] []
-------------
lea [a v] []

pn [a v b] []
-------------
lea [v b] []

<u>Example</u> :: We construct the conjunction [conj [pa pb] [p]] where

val[pa]=[lbx [q] [a]]
val[pb]=[ubx [q] [b]]
val[p]=[pn [q] [a b]]

The application specific axiom of the first kind for this conjunction states that if both APs [lbx [q] [a]] and [ubx [q] [b]] are computable then so is the conjunction [pn [q] [a b]], i.e.

lbx [q] [a]
ubx [q] [b]
------------
pn [q] [c d]

Equating the VAs of the output variables c and d with a and b, respectively, can be done in the application specific axioms of the second kind.

The application specific axioms of the second kind for this conjunction state that
if the conjunction [pn [q] [a b]] is computable then so are the APs [lbx [q] [c]]
and [ubx [q] [d]], i.e.

```
pn [q] [a b]                    pn [q] [a b]
------------                    lbx [q] [c]
lbx [q] [c]                     ------------
                                eqa [c a] []

pn [q] [a b]                    pn [q] [a b]
------------                    ubx [q] [d]
ubx [q] [d]                     ------------
                                eqa [d b] []
```

<u>Example</u> :: We construct the conjunction [conj [pa pb] [p]] where

```
val[pa]=[lbx [q] [a]]
val[pb]=[lea [a v] []]
val[p]=[pn [q v] []]
```

The application specific axiom of the first kind for this conjunction states that
if both APs [lbx [q] [a]] and [lea [a v] []] are computable then so is the
conjunction [pn [q v] []], i.e.

```
lbx [q] [a]
lea [a v] []
------------
pn [q v] []
```

The first application specific axiom of the second kind for this conjunction states
that if the conjunction [pn [q v] []] is computable then so is the AP
[lbx [q] [a]], i.e.

```
pn [q v] []
------------
lbx [q] [a]
```

Since cap[xb ya] is not the empty list we need to include [lbx [q] [a]] in the
premise of the second of the application specific axiom of the second kind to
conclude the computability of [lea [a v] []], i.e.

```
pn [q v] []
lbx [q] [a]
------------
lea [a v] []
```

<u>Notes</u> :: While introducing conjunctions will allow programs to be written in a more
concise form the decision to do so when using the proof assistant software VPC
should be balanced against the effort in constructing the application specific
axioms. This is not an issue with the software package RA where application
specific axioms for both disjunctions and conjunctions emerge as part of RA's
overall automated conjecturing process.

# 8    Machine numbers

## 8.1  Introduction

Here we will present some basic rules of machine numbers, type nat, under the constraints of the machine environment M[mach]. This will be the first part of an application of PECR that will be denoted by

nat[pname ax mach]

In the application nat[pname ax mach] we will deal exclusively with objects of type nat. Later we will extend this application to include objects of type arr and box. For the purposes of setting up the initializing data for the application nat[pname ax mach] in its entirety we define the list, tname, of types that will be used.

tname=[nat arr box]

*Table 18: Types of objects of the application* nat[pname ax mach]

| Integer label | Alphanumeric character label | Object |
|---|---|---|
| 1 | nat | machine number |
| 2 | arr | array of machine numbers |
| 3 | box | bounded regions of arrays of machine numbers |

## 8.2  Atomic programs

We will not explore arithmetic of machine numbers here. Only some order axioms will be presented. The type and relation checking programs are given by the following table. They are the first three elements of the list of APs for the application nat[pname ax mach].

*Table 19: Type and relation checking atomic programs*

| Integer label | Alphanumeric representation | Type and relation checks |
|---|---|---|
| 1 | [typen [a] []] | type[a nat] |
| 2 | [eqn [a b] []] | type[a nat], type[b nat], eqn[a b] |
| 3 | [lt [a b] []] | type[a nat], type[b nat], lt[a b] |

The list of constants, cst, of the application nat[pname ax mach] is given by

cst=[0 1 mnat]

where the elements 0 and 1 should be regarded as alphanumeric character strings that are assigned the fixed numerical values zero and one, respectively.

## 8.3  Axioms

*Equality*

```
Axiom axn1

typen [a] []
------------
eqn [a a] []
```

Equality is transitive, i.e.

```
eqn [a b] []
eqn [b c] []
------------
eqn [a c] []
```

This is not stated as an axiom because it follows immediately from the substitution rule.

Equality is also symmetric, i.e

```
eqn [a b] []
------------
eqn [b a] []
```

This follows from the IOT axiom, axiom axn1 and the substitution rule.

*Transitivity of inequality*

```
Axiom ord1

lt [a b] []
lt [b c] []
-----------
lt [a c] []
```

We include the following axioms, the first of which has an empty list premise.

| Axiom ord2a | Axiom ord2b |
|---|---|
| <br>-----------<br>lt [0 1] [] | lt [0 x] []<br>-----------<br>le [1 x] [] |

We also include the axiom of falsity

```
Axiom ord3

lt [a a] []
-----------
false
```

The following axioms state the upper and lower bounds of type nat objects.

| Axiom ord4a | Axiom ord4b |
|---|---|
| typen [a] []<br>------------<br>le [0 a] [] | typen [a] []<br>--------------<br>le [a mnat] [] |

Finally we have the axiom

| Axiom trich |
|---|
| typen [a] []<br>typen [b] []<br>--------------<br>trich [a b] [] |

The APs [le [a b] []] and [trich [a b] []] are disjunctions and are defined in the next section.

<u>Notes</u> :: Axiom ord3 has the HOP representation

```
-----------
flse [p] []
```

where p is a constant of type prgm under the HOP construct and has the fixed VA val[p]=[lt [a a] []], i.e. [lt [a a] []] is a constant of type prgm. Note that as a HOP construct the statement [flse [p] []] is a computable program while the ZOP, [lt [a a] []], is not computable for any VA of the input element, a.

## 8.4  Application specific disjunctions

We introduce the special disjunction programs that are associated with the terms le[a b] and trich[a b]. The term trich[a b] is defined if at least one of lt[b a] or le[a b] is defined. The APs that are associated with these terms are given in the following table.

*Table 20: Disjunctions*

| Integer label | Alphanumeric representation | Type and relation checks |
|---|---|---|
| 4 | [le [a b] []] | type[a nat], type[b nat], le[a b] |
| 5 | [trich [a b] []] | type[a nat], type[b nat], trich[a b] |

<u>Less than or equal</u>

The program [le [a b] []] is an AP that is constructed by the HOP

[disj [p q] [s]]

where p, q and s have the VAs

```
val[p]=[lt [a b] []]
val[q]=[eqn [a b] []]
val[s]=[le [a b] []]
```

The following axioms are included.

| Axiom le1 | Axiom le2 |
|---|---|
| lt [a b] [] <br> ----------- <br> le [a b] [] | eqn [a b] [] <br> ------------ <br> le [a b] [] |

These are the applications specific axioms of the disjunction [le [a b] []] as outlined in Section 7.3.

<u>Trichotomy</u>

Having constructed the disjunction [le [a b] []] we can now introduce the trichotomy program that is constructed by the HOP

```
[disj [p q] [s]]
```

where p, q and s have the VAs

```
val[p]=[lt [b a] []]
val[q]=[le [a b] []]
val[s]=[trich [a b] []]
```

<u>Notes</u> :: As outlined in Section 7.3 we can write the application specific axioms for the disjunction [trich [a b] []] as

```
lt [b a] []                    le [a b] []
--------------                 --------------
trich [a b] []                 trich [a b] []
```

However these need not be included as axioms because they follow from the IOT axiom applied to both input variables and axiom trich.

# 9 Discrete boxes

## 9.1 Vectors and arrays

A vector is a list of unit rank whose elements are all machine numbers, type nat. Vectors are assigned the type vec. The term

```
type[u vec[n]]
```

states that u is a vector of length n.

An array of type arr is a multidimensional list where each atomic element of the array is type nat. The dimension list of an array, u, is given by a vector, d,

```
type[u arr]
d=dim[u]
type[d vec[m]]
```

where m is the rank of the array. A vector can be regarded as an array of unit rank. We use the parameter dependent type arr[d] and write

```
type[m nat1]
type[d vec[m]]
type[u arr[d]]
```

to indicate that u is an array of dimension list d and rank m.

For two arrays, a and b, with the same dimension list we write

```
dim[a]=dim[b]
```

<u>Array equality and inequality</u>

When dealing with arrays we introduce the index vector. The index vector i, indv[i d], is defined by

```
type[d vec[m]]
indv[i d]
type[i vec[m]]
i=[i[1] i[2] ... i[m]]
j=1,2,...,m
   le[1 i[j]]
   le[i[j] d[j]]
```

The elements of an array, u, of dimensional list, d, can be expressed as functions of the index vector.

```
type[u arr[d]]
type[d vec[m]]
forall i indv[i d]
   u[i]=u[i[1]][i[2]]...[i[m]]
```

Two arrays, a and b, of type arr are said to be equal, and written

```
eqa[a b]
```

if they have the same dimension list and the VA of each corresponding atomic
element is equal. This can be expressed in terms of the index vector

```
eqa[a b]
type[a arr[d]]
type[b arr[d]]
type[d vec[m]]
forall i indv[i d]
   eqn[a[i] b[i]]
```

For two arrays a and b of type arr with the same dimension list we define
inequality by the term

```
lta[a b]
```

to mean that the VA of each atomic element of a is less than the VA of the
corresponding atomic element of b. In terms of the index vector

```
lta[a b]
type[a arr[d]]
type[b arr[d]]
type[d vec[m]]
forall i indv[i d]
   lt[a[i] b[i]]
```

For two arrays a and b of type arr with the same dimensions we define the term

```
lea[a b]
```

to mean that the VA of each atomic element of a is less than or equal to the VA of
each corresponding atomic element of b. In terms of the index vector

```
lea[a b]
type[a arr[d]]
type[b arr[d]]
type[d vec[m]]
forall i indv[i d]
   le[a[i] b[i]]
```

While scalar inequality le[a b] can be defined as a disjunction its array counter
part, lea[a b], cannot be constructed as a disjunction.

## 9.2  Atomic programs for arrays

We will not consider in any detail operations of array addition and products. They
can be examined in a similar manner to that of integer arrays of the earlier works
[9] and [27]-[29]. The following APs and axioms will be sufficient for our present
purposes.

*Table 21: Type and relation checking atomic programs*

| Integer label | Alphanumeric representation | Type and relation checks |
|---|---|---|
| 6 | `[typea [a] []]` | `type[a arr]` |
| 7 | `[eqa [a b] []]` | `type[a arr], type[b arr]` `dim[a]=dim[b], eqa[a b]` |
| 8 | `[lta [a b] []]` | `type[a arr], type[b arr]` `dim[a]=dim[b], lta[a b]` |
| 9 | `[lea [a b] []]` | `type[a arr], type[b arr]` `dim[a]=dim[b], lea[a b]` |

*Equality axiom*

```
Axiom axa1

typea [a] []
------------
eqa [a a] []
```

Transitivity of array equality

```
eqa [a b] []
eqa [b c] []
------------
eqa [a c] []
```

is not stated as an axiom because it follows immediately from the substitution rule. Array equality is also symmetric, i.e.

```
eqa [a b] []
------------
eqa [b a] []
```

This is not stated as an axiom because it follows from the IOT axiom, axiom axa1 and the substitution rule.

*Transitivity of inequality*

```
Axiom axa2

lta [a b] []
lta [b c] []
------------
lta [a c] []
```

The following axioms involve non-strict inequalities.

| Axiom axa3a | Axiom axa3b | Axiom axa3c | Axiom axa3d |
|---|---|---|---|
| eqa [a b] []<br>------------<br>lea [a b] [] | lta [a b] []<br>------------<br>lea [a b] [] | lea [a b] []<br>lea [b c] []<br>------------<br>lea [a c] [] | lea [a b] []<br>lea [b a] []<br>------------<br>eqa [b a] [] |

## 9.3  Intervals and boxes

A discrete interval can be represented by the ordered list of machine numbers, type nat,

[a a+1 ... a+n]

where a and a+n, for some n, are the lower and upper bound, respectively, of the interval.

A discrete interval can be extended to a multi-dimensional discrete box [a b], where the lower and upper bounds, a and b, are arrays of type arr. We write boxes as a two element list

[a b]
type[a arr[d]]
type[b arr[d]]
type[d vec[m]]
lea[a b]

A box, p=[a b], is assigned the generic type box to distinguish it from a standard two element list. We define the upper and lower bounds by the terms

type[p box]
p=[a b]
a=lbx[p]
b=ubx[p]

The box, p, is assigned the parameter dependent type box[d] such that d is the dimension list of its upper and lower bounds. We write

type[p box[d]]
type[lbx[p] arr[d]]
type[ubx[p] arr[d]]

for some dimension list, d, type[d vec[m]], of the box bounds. Note that a box p=[a b], type[p box[d]], is represented by an array of dimension list

d'=conclst[[2] d]
subtype[box[d] arr[d']]

If u, is an array, type[u arr[d]], such that, lea[a u] lea[u b], then we say that u is an element of the box p=[a b] and write

eltbx[u p]

## 9.4  Atomic programs of boxes

We introduce the following APs for boxes. There are four type and relation checking APs and three value and type assignment APs.

*Table 22: Type and relation checking APs for boxes*

| Integer label | Alphanumeric representation | Type and relation checks |
|---|---|---|
| 10 | [typebx [p] []] | type[p box] |
| 11 | [eqbx [p q] []] | type[p box], type[q box] eqa[lbx[p] lbx[q]] eqa[ubx[p] ubx[q]] |
| 12 | [eltbx [u p] []] | type[p box], type[u arr] eltbx[u p] |
| 13 | [subbx [q p] []] | type[q box], type[p box] lea[lbx[p] lbx[q]] lea[ubx[q] ubx[p]] |

*Table 23: Value and type assignment APs for boxes*

| Integer label | Alphanumeric representation | Type and relation checks | Value and type assignment |
|---|---|---|---|
| 14 | [lbx [p] [a]] | type[p box] | a:=lbx[p] type[a arr] |
| 15 | [ubx [p] [b] | type[p box] | b:=ubx[p] type[b arr] |
| 16 | [box [a b] [p]] | type[a arr], type[b arr], dim[a]=dim[b], lea[a b] | p:=[a b] type[p box[dim[a]]] |

The program [box [a b] [p]] constructs the box p=[a b], type[p box]. We use the same name of the program as the assigned type name of its output variable.

## 9.5  Box axioms

*Reflexivity*

| Axiom bx1 |
|---|
| typebx [p] [] -------------- eqbx [p p] [] |

Transitivity of box equality follows immediately from the substitution rule. Reflexivity of box equality can be derived as will be demonstrated in the next section.

*Box bounds*

| Axiom bx2a | Axiom bx2b | Axiom bx2c | Axiom bx2d |
|---|---|---|---|
| ```
typebx [p] []
--------------
lbx [p] [a]
``` | ```
typebx [p] []
--------------
ubx [p] [b]
``` | ```
lbx [p] [a]
ubx [p] [b]
--------------
lea [a b] []
``` | ```
lbx [p] [a]
ubx [p] [b]
lbx [q] [c]
ubx [q] [d]
eqa [c a] []
eqa [d b] []
--------------
eqbx [q p] []
``` |

*Box elements*

| Axiom bx3a | Axiom bx3b | Axiom bx3c |
|---|---|---|
| ```
lbx [p] [a]
eltbx [v p] []
--------------
lea [a v] []
``` | ```
ubx [p] [b]
eltbx [v p] []
--------------
lea [v b] []
``` | ```
lbx [p] [a]
ubx [p] [b]
lea [a v] []
lea [v b] []
--------------
eltbx [v p] []
``` |

*Subbox*

| Axiom bx4a | Axiom bx4b | Axiom bx4c |
|---|---|---|
| ```
lbx [p] [a]
lbx [q] [c]
subbx [q p] []
--------------
lea [a c] []
``` | ```
ubx [p] [b]
ubx [q] [d]
subbx [q p] []
--------------
lea [d b] []
``` | ```
lbx [p] [a]
lbx [q] [c]
lea [a c] []
ubx [p] [b]
ubx [q] [d]
lea [d b] []
--------------
subbx [q p] []
``` |

*Box construction*

| Axiom bx5a | Axiom bx5b | Axiom bx5c |
|---|---|---|
| ```
lea [a b] []
--------------
box [a b] [p]
``` | ```
box [a b] [p]
lbx [p] [c]
--------------
eqa [c a] []
``` | ```
box [a b] [p]
ubx [p] [c]
--------------
eqa [c b] []
``` |

## 9.6  Basic properties of boxes

We will now check to see if the axioms bx1-bx5 are strong enough to allow us to derive some basic properties of boxes. The proofs shown here were generated by RA's automated theorem prover and validated by VPC's proof checking component. They are
presented in the output format used by VPC.

Theorems are labeled by the alphanumeric character string thm followed by a number.
We start by showing that subboxes are reflexive.

thm1

```
typebx [p] []
--------------
subbx [p p] []
```

proof

```
 1  typebx [p] []
 2  eqbx [p p] []                  bx1 [1]
 3  lbx [p] [a]                    bx2a [1]
 4  ubx [p] [b]                    bx2b [1]
 5  lbx [p] [c]                    sr1 [3 2]
 6  ubx [p] [d]                    sr1 [4 2]
 7  eqa [d b] []                   sr2 [4 2 6]
 8  eqa [a c] []                   sr2 [5 2 3]
 9  lea [d b] []                   axa3b [7]
10  lea [a c] []                   axa3b [8]
11  subbx [p p] []                 bx4c [3 5 10 4 6 9]
```

The proof of thm1 makes use of repeated programs with nonempty output lists where,
by necessity, the output variable labels differ (statements 3 and 5 and statements
4 and 6). These can be associated to the first part of the identity rule that
follows from the application of the first part of the substitution rule, sr1, using
the equality program [eqbx [p p] []] (Section 5.3). The output variables for each
pair of assignment programs are subsequently equated by the second part of the
substitution rule, sr2.

Subboxes are also transitive.

thm2

```
subbx [p q] []
subbx [q r] []
--------------
subbx [p r] []
```

proof

```
 1  subbx [p q] []
 2  subbx [q r] []
 3  typebx [p] []                  iot [1]
 4  typebx [q] []                  iot [1]
 5  typebx [r] []                  iot [2]
 6  lbx [p] [a]                    bx2a [3]
 7  lbx [q] [b]                    bx2a [4]
 8  lbx [r] [c]                    bx2a [5]
 9  ubx [p] [d]                    bx2b [3]
10  ubx [q] [e]                    bx2b [4]
11  ubx [r] [f]                    bx2b [5]
12  lea [b a] []                   bx4a [7 6 1]
```

```
13  lea [c b] []                        bx4a [8 7 2]
14  lea [d e] []                        bx4b [10 9 1]
15  lea [e f] []                        bx4b [11 10 2]
16  lea [c a] []                        axa3c [13 12]
17  lea [d f] []                        axa3c [14 15]
18  subbx [p r] []                      bx4c [8 6 11 9 17]
```

Theorem thm3 associates subboxes with box equality.

thm3

```
subbx [p q] []
subbx [q p] []
--------------
eqbx [q p] []
```

proof

```
 1  subbx [p q] []
 2  subbx [q p] []
 3  typebx [p] []                       iot [1]
 4  typebx [q] []                       iot [1]
 5  lbx [p] [a]                         bx2a [3]
 6  lbx [q] [b]                         bx2a [4]
 7  ubx [p] [c]                         bx2b [3]
 8  ubx [q] [d]                         bx2b [4]
 9  lea [a b] []                        bx4a [5 6 2]
10  lea [b a] []                        bx4a [6 5 1]
11  lea [d c] []                        bx4b [7 8 2]
12  lea [c d] []                        bx4b [8 7 1]
13  eqa [b a] []                        axa3d [9 10]
14  eqa [d c] []                        axa3d [12 11]
15  eqbx [q p] []                       bx2d [5 7 6 8 13 14]
```

Theorem thm4 shows that an element, v, of a box, p, will also be an element of a box, q, that encloses the box p.

thm4

```
eltbx [v p] []
subbx [p q] []
--------------
eltbx [v q] []
```

proof

```
 1  eltbx [v p] []
 2  subbx [p q] []
 3  typebx [p] []                       iot [1]
 4  typebx [q] []                       iot [2]
 5  lbx [p] [a]                         bx2a [3]
 6  lbx [q] [b]                         bx2a [4]
 7  ubx [p] [c]                         bx2b [3]
 8  ubx [q] [d]                         bx2b [4]
 9  lea [a v] []                        bx3a [5 1]
10  lea [v c] []                        bx3b [7 1]
11  lea [b a] []                        bx4a [6 5 2]
```

```
12  lea [c d] []                        bx4b [8 7 2]
13  lea [v d] []                        axa3c [10 12]
14  lea [b v] []                        axa3c [11 9]
15  eltbx [v q] []                      bx3c [6 8 14 13]
```

Axiom bx1 states that box equality is reflexive. Box equality is also symmetric as the following demonstrates.

thm5

```
eqbx [p q] []
-------------
eqbx [q p] []
```

proof

```
  1  eqbx [p q] []
  2  typebx [p] []                      iot [1]
  3  eqbx [p p] []                      bx1 [2]
  4  eqbx [q p] []                      sr1 [3 1]
```

Transitivity of box equality follows immediately from the substitution rule.

The following associates box equality with subboxes.

thm6

```
eqbx [p q] []
--------------
subbx [p q] []
```

proof

```
  1  eqbx [p q] []
  2  typebx [q] []                      iot [1]
  3  subbx [q q] []                     thm1 [2]
  4  eqbx [q p] []                      thm5 [1]
  5  subbx [p q] []                     sr1 [3 4]
```

## 9.7  Dynamical systems

An understanding of the universe is acquired through the process of constructing models of it. Working under the hypothesis of a computable universe we have accepted that the laws of nature must be based on a discrete formulation. Computer models of real-world phenomena are built upon the laws that are the end product of rigorous program constructions based on CoL as implemented by the formal system PECR. While these computer models can serve as simulation tools they should be regarded as the ultimate expression of the laws that govern the application under consideration. This opposes the traditional approach of representing the laws of nature by mathematical equations.

Many applications of real-world phenomena come in the form of dynamical systems. Here we will examine the computability of *fully discrete dynamical systems*. The expression *fully discrete dynamical systems* is used to maintain a clear distinction from the expression *discrete dynamical systems* that is commonly used to describe the iteration of real-valued functions where only time is discrete.

Consider the function

v=f[u]

that maps the integer array u to the integer array v, where d=dim[u]=dim[v]. The domain of the function f[u] can be defined by a box p where the argument u of the function f[u] is an element of p.

The range of the function f[u] will be contained in a box q such that for any u, eltbx[u p], we have eltbx[f[u] q]. The range of f over a domain p is expressed by the term

range[f p]

A fully discrete dynamical system can be written in the abstract form

u[t+1]=f[u[t]],    t=0,1,...

where u[t], type[u[t] arr[d]], t=0,1,.... Let the term

u[n]=itf[u[0] n]

represent the value of u[n] after n iterations of the function f[:] with the starting value u[0].

Suppose that

[f [u] [v]]

is the program that is a realization of the function represented by the term v=f[u], where u and v are arrays with the same dimension list, d. In the language of functional programs we can write the dynamical system as a vertical list

f [u[0]] [u[1]]
f [u[1]] [u[2]]
.
.
.
f [u[n-1]] [u[n]]

On a machine each u[t], type[u[t] arr[d]], t=0,1,...,n, will be represented by a unique variable label. For large n it is more practical to introduce the iteration program

[itf [u n] [w]]

whose internal code is constructed by some imperative language.

*Algorithm 5: Iteration program* [itf [u n] [w]]

```
le[0 n]
w:=u
do t=1 to n
   w:=f[w]
enddo
```

When n=0 the do loop in the program [itf [u n] [w]] is bypassed and the array w, eqa[w u], is returned as output. Algorithm 5 represents a pseudo-code of some imperative language so the integer scalar t is defined internally and discarded from memory storage when the execution of [itf [u n] [w]] is completed.

The following APs define a specific application of a fully discrete dynamical system based on some function f[:].

*Table 24: Atomic programs associated with the application specific function* f[:]

| Integer label | Alphanumeric representation | Type checks | Value and type assignment |
|---|---|---|---|
| 17 | [f [u] [v]] | type[u arr] | v:=f[u]<br>type[v arr]<br>dim[v]=dim[u] |
| 18 | [itf [u n] [w]] | type[u arr]<br>type[n nat] | w:=itf[u n]<br>type[w arr]<br>dim[w]=dim[u] |
| 19 | [bndf [p] [q]] | type[p box] | q:=[a b]<br>type[q box]<br>lea[a lbx[range[f p]]]<br>lea[ubx[range[f p]] b] |

The AP [bndf [p] [q]] constructs a box q that contains range[f p]. Here we regard [f [u] [v]] as an AP. It is possible that [f [u] [v]] can be represented as a program list of APs. In such a case we can construct [f [u] [v]] recursively as a conjunction, type cnj (Section 6.4). The properties of [f [u] [v]] can be conveyed to the machine by way of application specific axioms of conjunctions as outlined in Section 7.4.

By construction the iteration assignment program [itf [u n] [w]] satisfies the following axioms.

| Axiom itf1 |
|---|
| itf [u 0] [w]<br>-------------<br>eqa [w u] [] |

This axiom reflects the property that the do-loop in Algorithm 5 is not activated when n=0 and the assignment w:=u is returned as output.

The iteration assignment program also obeys

| Axiom itf2a | Axiom itf2b | Axiom itf2c |
|---|---|---|
| f [u] [v]<br>-------------<br>itf [u 1] [w] | itf [u 1] [w]<br>-------------<br>f [u] [v] | itf [u 1] [w]<br>f [u] [v]<br>-------------<br>eqa [v w] [] |

Our main objective here is to establish that given a discrete box, type[p box[d]], if u, type[u arr[d]], is an element of the box p then the program [itf [u n] [w]] will be computable for any n, type[n nat]. We can expect that the computability of the program [itf [u n] [w]] will be guaranteed if the range, range[f p], is
contained in the box p.

It should be noted that by definition of the discrete function f[u] in a machine
environment M[mach] the elements of the array u can only take on a finite number of
assigned values over the box p.

We start with the box p that defines the domain of the function f[:]. Given
type[p box[d]] the user supplied program

[bndf [p] [q]]

attempts to construct q, type[q box[d]], such that q is a sufficiently tight box
that contains range[f p]. The internal algorithm of [bndf [p] [q]] is not directly
visible to the machine so we must supply additional application specific axioms
that inform the machine of the rules associated with the methods employed to obtain
the bounds of the box q. These bounds will be dependent on the bounds of the box p
and the internal algorithm of the program [f [u] [v]]. Whatever method we choose to
construct q from p it must be soundly based such that the following axioms hold.

| Axiom bndfa | Axiom bndfb |
|---|---|
| bndf [p] [q]<br>eltbx [u p] []<br>--------------<br>f [u] [v] | bndf [p] [q]<br>eltbx [u p] []<br>f [u] [v]<br>--------------<br>eltbx [v q] [] |

Axiom bndfa states that if [bndf [p] [q]] is computable and u is an element of the
box p then the evaluation v of [f [u] [v]] exists.

Axiom bndfb states that given the box p, if u is an element of p and v is obtained
from the evaluation [f [u] [v]] then v is an element of the box q. This is
equivalent to the statement that q will be a box that contains range[f p].

Axioms bndfa and bndfb will be satisfied by any program [f [u] [v]] that is
computable over its domain, p. However computability of [f [u] [v]] for all
elements u of p will not be sufficient to establish computability of its associated
iteration program [itf [u n] [w]].

Given the box p, the first task is to find a suitable box q such that q is a
sufficiently tight box containing range[f p]. If in addition we can construct q
such that q is contained in p then we can apply the following axiom of
computability.

| Axiom axc |
|---|
| bndf [p] [q]<br>subbx [q p] []<br>eltbx [u p] []<br>le [0 n] []<br>--------------<br>itf [u n] [w] |

Axiom axc states that given the box p, if the box q contains range[f p] and q is contained in p then for any element u of p the iteration program [itf [u n] [w]] will be computable for any n, type[n nat].

<u>Notes</u> ::

1. In practice it is often desirable to output the VAs of w of the iteration program [itf [u n] [w]] at regular iteration intervals. This is easily accomplished by inserting a write statement within the do loop of Algorithm 5 that prints intermediate VAs of the array w. In the interests of brevity this has not been included in the do loop because it does not have any direct relevance to the computability properties.

2. In some applications it may not be obvious that the internal algorithm of [bndf [p] [q]] satisfies axioms bndfa-bndfb. In such cases it is best to establish bndfa-bndfb as theorems from a more fundamental application. To this end one must provide application specific axioms that inform the machine of the rules that describe the internal algorithms of the programs [f [u] [v]] and [bndf [p] [q]].

3. Axiom axc will always hold in principle but will only be of practical use if one is able to find sufficiently tight bounds of the function f[u]. There may be dynamical systems where sufficiently tight bounds for f[u] will always be elusive. For such cases other methods of establishing computability need to be explored.

4. There are many applications where the bounds on the function f[u] are explicitly built into the internal algorithm that defines the function itself. Of special interest here are dynamical systems that are based on maps on configuration states of arrays as discussed in Section 1.4.

## 9.8 Cycles

In mathematics we may have a sequence of statements, P[n], that depend on the natural number n. We have by the principle of induction :: if P[0] is true and P[n] implies P[n+1] then P[n] is true for all natural numbers n.

Axiom axc states that a fully discrete dynamical system will be computable for any n such that le[0 n]. Since we are working on a real-world machine, M[mach], it follows that n cannot exceed the maximum machine number, mnat. At first glance we might suspect that, by analogy, axiom axc is weaker than a computability argument based on the principle of induction. To address this issue we need to make a few observations.

First of all, axiom axc is valid on any machine, M[mach]. This suggests that it is possible to increase the upper bound of the parameter n by working on a larger machine. But there is an important limitation here in that we are only interested in real-world machines.

The program

[f [u] [v]]

is the realization of the function represented by the term v=f[u] such that

d=dim[u]=dim[v]
type[d vec[m]]
eltbx[u p]
type[p box[d]]

A fully discrete dynamical system can be represented in the abstract form

u[t+1]=f[u[t]],  t=0,1,...

where u[t], type[u[t] arr[d]], t=0,1,....

According to axiom axc the dynamical system will be computable on a machine M[mach] if range[f p] is contained in the box p. Since u is an integer array contained in a bounded and discrete box p there must exist a finite number

nu[p]

of possible unique VAs of the array u in the box p. If range[f p] is contained in the box p it follows that the fully discrete dynamical system must eventually enter a cycle at some discrete time tcyc such that

le[tcyc nu[p]]

A cycle can be defined by

f[u[t+pcyc]]=f[u[t]],   le[tcyc t]

for some pcyc, type[pcyc nat1]. The parameter pcyc is called the *period* of the cycle. We must also have

le[pcyc nu[p]]

A cycle of period pcyc=1 is called a *fixed point*.

The problem with the above abstractions is that the parameters ncyc and/or pcyc may sometimes be so large that a cycle will not be realizable on a specific machine M[mach], at least in the sense of a feasible computation. It will often be the case that the only way to determine the values of tcyc and pcyc is to run the iteration program [itf [u n] [w]] for sufficiently large n in the hope of detecting a cycle. For such a task it is usually essential that a write statement be embedded within the do loop of Algorithm 5 so that assigned values of w can be accessed at regular iteration intervals for analysis (see first item of the notes in Section 9.7).

There have been extensive studies of dynamical systems based on real-valued functions where the time is either discrete or continuous. Very little has been done on fully discrete dynamical systems. The ideas presented here are only meant to be preliminary but they already indicate that the properties of fully discrete dynamical systems will differ significantly from those of dynamical systems based on real-valued functions. Further exploration into this topic is needed.

Notes :: An example of a one-dimensional fully discrete dynamical system by way of the discrete tent map has been briefly explored in Chapter 7 of [29].

# 10  Programs as integer arrays

## 10.1      Introduction

Proofs in mathematics are constructed using a mixture of natural language and symbols. Under this convention we sometimes lose sight of the fact that the construction of proofs is essentially just another application of computation[1]. In this chapter we will introduce an additional simplification to our formal language that will allow us to gain more insight into the fundamentals of proof construction as a computation.

Up until now we have expressed most of the labels of the elements of lists by alphanumeric character strings but we shall see that integers can serve just as well as labels. In practice all objects that are represented by strings of symbols are encoded into binary strings within the computer. This raises the question as to whether there are any benefits in translating our formal language directly into binary strings. The down side of this is that the representation of objects by arrays of binary strings can be rather lengthy to write down. It is more convenient to use the shorthand notation of the decimal representation of integers with the understanding that decimal and binary numbers can be readily converted from one to the other.

Alphanumeric representations of objects can become cumbersome when dealing with a large number of labels. Integer labels have the property of conciseness and can be efficiently selected from a large ordered list. This is advantageous from a 'book keeping' point of view. Once a systematic procedure for integer labeling is set up the essential components of the computational operations of maps associated with proof constructions come into light.

## 10.2      Integer labeling of programs

We start by defining the environment within which we will work. The element mlst of the machine parameters list, mach, will be regarded as a list

mlst=[nprem npmax nx ny]

where

nprem = maximum length of any premise program list
npmax = maximum length of any program list
nx = maximum length of the input list of any AP
ny = maximum length of the output list of any AP

We can regard the first 2 parameters, nprem and npmax, as dictating the depth in which a rational agent wishes to explore an application.

We have been representing APs using the lists

pname=[pname[1] pname[2] ... pname[nap]]
var=[var[1] var[2] ... var[nvar]]
cst=[cst[1] cst[2] ... cst[ncst]]

---

1   One notable figure who contests this is Roger Penrose who argues that the human mind has the ability to recognize truths by a process that cannot be explained by computation alone [30].

The lists pname and cst are application specific. For each application we also include the list of type names

tname=[tname[1] tname[2] ... tname[ntype]]

Names/labels of programs, I/O variables, constants and types can be represented by integers. Integer labels of program names and types of each application appear in the first columns of their relevant tables presented in the previous chapters. We use the convention that for any list of atomic elements

a=[a[1] a[2] ... a[na]]

set

a[i]=i, i=1,2,...,na

For instance upon output we can convert each integer element label of, var, to a letter in the corresponding order of the alphabet of the natural language, a,b,...,z. We use the convention $var[1] \longmapsto a$, $var[2] \longmapsto b$, ... , $var[26] \longmapsto z$. If more variables are needed the sequence of letters are repeated with an appended number, a1,b1,...,z1, and so on.

A program, p, of list length n can be represented by

p=[p[1] p[2] ... p[n]]

where each p[i], i=1,2,...,n, is an AP that has the list structure

p[i]=[pn[i] x[i] y[i]]

where pn[i] is the program name of p[i], x[i] is the input list and y[i] is the output list. The program p=[p[1] p[2] ... p[n]] is translated into the integer array as follows.

type[p prgm[n]]
subtype[prgm[n] arr[d]]
type[d vec[2]]
d=[n 1+nx+ny]
i=1,2,...,n
    type[pn[i] nat1]
    type[x[i] vec[nx]]
    type[y[i] vec[ny]]
    type[p[i] vec[1+nx+ny]]
    p[i]=chain[pn[i] x[i] y[i]]

The scalar pn[i] is treated as a singleton under the action of chain[:]. Recall that the action of chain[:] removes the outermost brackets of lists enclosing its elements (Section 2.4).

Some APs will have I/O lists that have shorter lengths than the maximum lengths nx and ny. We maintain the condition type[x[i] vec[nx]] and type[y[i] vec[ny]] and define an empty element of these lists as a *null variable*.

Each element, p[i], of the program list p will be an AP. The I/O lists of each p[i], i=1,2,...,n, have the expanded form

```
x[i]=[x[i][1] x[i][2] ... x[i][nx]]
```

```
y[i]=[y[i][1] y[i][2] ... y[i][ny]]
```

Each element of the I/O lists x[i] and y[i] is assigned the integer label under the following convention.

```
x[i][j] = {k,       le[1 k] le[k nvar], x[i][j] is associated with variable var[k]
          {nvar+m,  le[1 m] le[m ncst], x[i][j] is associated with constant cst[m]
          {0,       x[i][j] is a null variable
```

```
y[i][j] = {k,       le[1 k] le[k nvar], y[i][j] is associated with variable var[k]
          {0,       y[i][j] is a null variable
```

Here we are effectively appending the list of integer labels of the constants, cst, to the list of integer labels of variable labels, var, i.e. an integer label nvar+m, le[1 m] le[m ncst], is associated with a constant cst[m]. A label k, le[1 k] le[k nvar], is associated with the variable var[k].

Using this format we can now write a program p, type[p prgm[n]], as an integer matrix

```
[pn[1] x[1][1] x[1][2] ... x[1][nx] y[1][1] y[1][2] ... y[1][ny]]
[pn[2] x[2][1] x[2][2] ... x[2][nx] y[2][1] y[2][2] ... y[2][ny]]
.                                                              .
.                                                              .
.                                                              .
[pn[n] x[n][1] x[n][2] ... x[n][nx] y[n][1] y[n][2] ... y[n][ny]]
```

The *I/0 matrix* of a program is obtained by removing the first column of this array.

VPC and RA will automatically allocate the length, nvar, of the variable labels list var. Within VPC and RA programs are represented largely using the integer format convention described here. Under this convention VPC and RA perceive the process of program constructions as maps over integer arrays. Programs are translated to and from the integer and alphanumeric formats only for human readability of certain I/O data files.

<u>Example</u> :: In alphanumeric format the derivation program is given by

```
ext [q c] []
sub [q p] []
conc [p c] [s]
```

The complete list of distinct I/O variable labels that appear in this program is given by

```
[q c p s]
```

The integer representation of this list is

```
[17 3 16 19]
```

The derivation program can be encoded into the integer matrix

```
[6 17  3  0]
[4 17 16  0]
[8 16  3 19]
```

where we are using nx=2 and ny=1. The first column contains the integer labels of
the program names as presented in the first column of the tables of APs in Section
3.7. The remaining elements of each row are the integer labels of the variables.
Null variables are set to 0.

The I/O matrix for this program is obtained by removing the first column and is
given by

```
[17  3  0]
[17 16  0]
[16  3 19]
```

The integer label of each non null variable is arbitrary provided it is bounded
above by nvar and preserves the same configuration of repetitions.

<u>Notes</u> :: The integer labels of the elements of the I/O lists do not give any
information of the actual VAs attached to them. VAs require an additional map that
sends each integer I/O element label to an object that represents the actual VA of
that element. Objects that represent VAs can also be translated from an
alphanumeric format to integer scalars and arrays.

## 10.3      Decomposition of I/0 matrices

An I/O matrix of a program can be decomposed into matrices whose only nonzero
elements correspond to a distinct I/O element label. The complete list of distinct
I/O element labels that appear in a program, p, is given by

lio[p]=unique[conclst[inp[p] outp[p]]

The unique function is applied to remove repetitions of labels. The input list,
inp[p], can contain labels taken from the list var as well as cst while the output
list, outp[p], can only contain labels taken from var.

Let mio be the I/O matrix of the program p, type[p prgm[n]]. For each I/O element
label

lio[p][k]=elt[lio[p] k]

we construct the matrices u[k], k=1,2,...,len[lio[p]], as follows.

```
type[mio arr[d]]
d=[n nx+ny]
for k=1,2,...,len[lio[p]]
    type[u[k] arr[d]]
    for i=1,2,...,n
        for j=1,2,...,nx+ny
            u[k][i][j]={lio[p][k],   mio[i][j]=lio[p][k]
                      {0,             otherwise
```

Each matrix, u[k], in this decomposition of the I/O matrix, mio, of the program p
is assigned the subtype

type[u[k] dmio[p k]]

For type[p prgm[n]] we have subtype[dmio[p k] arr[d]], d=[n nx+ny]. The integer k in the parameter dependent type dmio[p k] associates the matrix u[k] with the I/O element label lio[p][k].

The original I/O matrix is obtained by the standard matrix sum

mio=u[1]+u[2]+...+u[m]

where m=len[lio[p]].

A matrix u[k], type[u dmio[p k]], that contains more than one nonzero label or an element that is bound to a constant is said to be an *I/O binding matrix*.

Binary matrices can be employed as templates for matrices of type dmio. More generally a binary array is an integer array whose elements are 0 or 1. Binary arrays will be assigned the type barr where

subtype[barr arr]

For each u[k], type[u[k] dmio[p k]], k=1,2,...,len[lio[p]], of the decomposition of the I/O matrix, mio, of a program p, type[p prgm[n]], there exists a binary matrix, b[k], such that

d=[n nx+ny]
**for** k=1,2,...,len[lio[p]]
   type[u[k] dmio[p k]]
   type[b[k] barr[d]]
   u[k]=lio[p][k]∗b[k]

where the symbol ∗ represents the standard scalar product of the scalar lio[p][k] and the matrix b[k].

A binary matrix, b[k], that is a template for a matrix u[k], type[u[k] dmio[p k]], is assigned the type

type[b[k] bmio[p k]]

<u>Example</u> :: Consider axiom axc of the application nat[pname ax mach] given in Section 9.7 as the IEP

bndf [p] [q]
subbx [q p] []
eltbx [u p] []
le [0 n] []
itf [u n] [w]

where the dashed line that separates the premise program from the conclusion program has been omitted.

The list of distinct I/O element labels for this program is given by the alphanumeric representation

[p q u 0 n w]

The fourth element is the constant 0 appearing as the first element, cst[1], of the list of constants, cst, presented in Section 8.2 and is treated as an alphanumeric string. In the integer format the list of distinct I/O element labels is given by

[16 17 21 1* 14 23]

where we have used the shorthand notation

1*=nvar+1

to represent the integer label of the constant cst[1].

The program can be encoded into the integer array

```
[19 16  0 17]
[13 17 16  0]
[12 21 16  0]
[ 4 1* 14  0]
[18 21 14 23]
```

where we are using nx=2 and ny=1. The first column contains the integer labels of the program names as presented in the first columns of the tables of APs, Tables 20, 22 and 24. The remaining elements of each row are the integer labels of the I/O elements. Null variables are set to 0.

The I/O matrix, mio, is obtained by removing the first column of the integer matrix of the program to obtain

```
[16  0 17]
[17 16  0]
[21 16  0]
[1* 14  0]
[21 14 23]
```

The integer label of each non null variable is arbitrary provided it is bounded above by nvar and preserves the same configuration of repetitions.

The matrices u[k], k=1,2,...,6, of the decomposition of the I/O matrix, mio, are given by

```
[16  0  0]   [ 0  0 17]   [ 0  0  0]   [ 0  0  0]   [0  0  0]   [0  0  0]
[ 0 16  0]   [17  0  0]   [ 0  0  0]   [ 0  0  0]   [0  0  0]   [0  0  0]
[ 0 16  0]   [ 0  0  0]   [21  0  0]   [ 0  0  0]   [0  0  0]   [0  0  0]
[ 0  0  0]   [ 0  0  0]   [ 0  0  0]   [1*  0  0]   [0 14  0]   [0  0  0]
[ 0  0  0]   [ 0  0  0]   [21  0  0]   [ 0  0  0]   [0 14  0]   [0  0 23]
```

The first, second, third and fifth matrices have repeated variable labels and hence are I/O binding matrices. The fourth matrix does not contain any repetitions of labels but contains a label that is bound to a constant so it is also an I/O binding matrix. The sixth matrix is not an I/O binding matrix because it has no repetitions of variable labels or labels bound to a constant. The original I/O matrix is obtained by applying a standard matrix sum to these six matrices. The binary templates, b[k], of each matrix u[k] are obtained by simply replacing all nonzero elements with 1.

## 10.4      AND/OR logic gates

Let u and v be two binary arrays of the same dimensions. The array w denoted by the
term

w=and[u v]

is a binary array defined by

```
type[u barr[d]]
type[v barr[d]]
type[w barr[d]]
type[d vec[m]]
w=and[u v]
indv[i d]
i=[i[1] i[2] ... i[m]]
forall i indv[i d]
   w[i]={1,   u[i]=v[i]=1
        {0,   otherwise
```

The operator and[u v] acts as an AND logic gate for each element pair u[i] and
v[i], indv[i d].

The array w denoted by the term

w=or[u v]

is a binary array defined by

```
type[u barr[d]]
type[v barr[d]]
type[w barr[d]]
type[d vec[m]]
w=or[u v]
indv[i d]
i=[i[1] i[2] ... i[m]]
forall i indv[i d]
   w[i]={0,   u[i]=v[i]=0
        {1,   otherwise
```

The operator or[u v] acts as an OR logic gate for each element pair u[i] and v[i],
indv[i d].

Example ::

1. The and/or operators applied to the 3x3 matrices

u=[1 1 0]                                    v=[0 1 0]
  [0 1 0]                                      [1 1 0]
  [1 0 0]                                      [1 0 1]

yield

and[u v]=[0 1 0]                    or[u v]=[1 1 0]
         [0 1 0]                            [1 1 0]
         [1 0 0]                            [1 0 1]

2. The and/or operators applied to the 3x4 matrices

```
u=[1 1 1 0]                              v=[0 1 0 0]
  [1 0 0 1]                                [0 1 1 0]
  [0 1 0 1]                                [0 0 0 1]
```

yield

```
and[u v]=[0 1 0 0]                       or[u v]=[1 1 1 0]
         [0 0 0 0]                               [1 1 1 1]
         [0 0 0 1]                               [0 1 0 1]
```

I/O equivalence

We can now define I/O equivalence in terms of binary templates of type dmio matrices under the action of the operator and[:].

*Table 25: I/O equivalence* ioeq[q p]

| Condition 1 | type[q prgm[m]]<br>type[p prgm[m]] |
|---|---|
| Condition 2 | **for** i=1,2,...,m<br>    in[pn_p[i] pname]<br>    in[pn_q[i] pname]<br>    pn_q[i]=pn_p[i] |
| Condition 3 | **forall** k in[lio[p][k] var]<br>    type[b[k] bmio[p k]]<br>    **exists** lio[q][j]<br>    type[a[j] bmio[q j]]<br>    and[a[j] b[k]]=b[k] |
| Condition 4 | **forall** k in[lio[p][k] cst]<br>    type[b[k] bmio[p k]]<br>    **exists** lio[q][j]=lio[p][k]<br>    type[a[j] bmio[q j]]<br>    and[a[j] b[k]]=b[k] |

where pn_q[i] and pn_p[i] are the labels of the program names of the subprograms q[i] and p[i], respectively.

Notes ::

1. In Condition 3 the binary array templates b[k] must be associated with I/O element labels of the program p that are variables while the binary array templates a[j] can be associated with any I/O element label of the program q. In Condition 4 the binary array templates b[k] and a[j] must be associated with the same constant.

2. The I/O lists of any program can be determined from the binary binding matrices alone. Consequently we can exclude type bmio[p k] and bmio[q j] binary matrices that are not templates of I/O binding matrices in Conditions 3 and 4.

## 10.5    Derivation maps

Having constructed the integer labeling convention of programs we are now in a position to view program constructions from the perspective of maps on integer matrices. Each row of a program matrix is a placeholder for an AP that is inserted into the matrix from top to bottom as the program construction proceeds.

In particular, the construction of IEPs can be viewed as maps on integer matrices. A premise of length n, le[n nprem], occupies the top n rows of the initial matrix. In the target array of this map the conclusion program is appended immediately below the rows that store the premise program. In this way we have defined the construction of an IEP of length n+1 as a map

map[arr[n nx+ny+1] arr[n+1 nx+ny+1]]

The premise of an IEP p, type[p prgm[n]], is represented by the integer matrix of type arr[n nx+ny+1].

*Premise matrix*

```
[pn[1] x[1][1] x[1][2] ... x[1][nx] y[1][1] y[1][2] ... y[1][ny]]
[pn[2] x[2][1] x[2][2] ... x[2][nx] y[2][1] y[2][2] ... y[2][ny]]
.                                                               .
.                                                               .
.                                                               .
[pn[n] x[n][1] x[n][2] ... x[n][nx] y[n][1] y[n][2] ... y[n][ny]]
```

The premise matrix of the IEP maps to an integer matrix of type arr[n+1 nx+ny+1] by appending the conclusion program p[n+1]=chain[pn[n+1] x[n+1] y[n+1]] to the premise matrix.

Example :: The alphanumeric character and integer representations of the program extension rule, per, can be written

```
sub [q p] []                        [4 17 16 19]
ext [q c] []                        [6 17  3  0]
conc [p c] [s]                      [8 16  3  0]
--------------                      ------------
aext [p c] []                       [9 16  3  0]
```

We can regard the program extension rule to represent a map on the integer matrices

```
[4 17 16  0]    ⟼   [4 17 16  0]
[6 17  3  0]         [6 17  3  0]
[8 16  3 19]         [8 16  3 19]
                     [9 16  3  0]
```

The labels in the first columns of these matrices are associated with program names and cannot be changed (Tables 8-10). The labels in the I/O part of the matrix can be changed provided that they are bounded above by nvar and maintain the same configurations of repeated labels.

# 10.6      Proof programs

In our original representations of programs the second condition of a PE states that the input list of the conclusion program can contain labels associated with constants but cannot introduce variable labels that are not contained in the I/O lists of the premise program (Section 4.2). In our integer representation of programs we can restate the second condition of a PE, type[c ext[p]], as the requirement that each element of the input list of the conclusion program c must belong to an I/O binding matrix of the program s=conc[p c] or to an I/O binding matrix associated with a constant. In order that the I/O dependency conditions are fully satisfied each element of output list of c must not belong to any binding matrix of s.

Each step of a proof construction in VPC involves the generation of all possible PEs of the current proof program. This is done by extracting all sublists from the current proof program that are I/O equivalent to the premises of currently known IEPs. From this collection of PEs is selected a new statement of the proof. The whole process is repeated until a final statement is accepted as the conclusion to the theorem. The following example presents the alphanumeric and integer representations of a completed proof.

Example :: The proof program of the theorem

thm2

subbx [p q] []
subbx [q r] []
--------------
subbx [p r] []

of the application nat[pname ax mach] (Section 9.6) has the alphanumeric character and integer matrix representations given in Tables 26 and 27, respectively.

*Table 26: Alphanumeric character representation of the proof program of theorem* thm2 *of the application* nat[pname ax mach]

```
 1  subbx [p q] []
 2  subbx [q r] []
 3  typebx [p] []              iot [1]
 4  typebx [q] []              iot [1]
 5  typebx [r] []              iot [2]
 6  lbx [p] [a]                bx2a [3]
 7  lbx [q] [b]                bx2a [4]
 8  lbx [r] [c]                bx2a [5]
 9  ubx [p] [d]                bx2b [3]
10  ubx [q] [e]                bx2b [4]
11  ubx [r] [f]                bx2b [5]
12  lea [b a] []               bx4a [7 6 1]
13  lea [c b] []               bx4a [8 7 2]
14  lea [d e] []               bx4b [10 9 1]
15  lea [e f] []               bx4b [11 10 2]
16  lea [c a] []               axa3c [13 12]
17  lea [d f] []               axa3c [14 15]
18  subbx [p r] []             bx4c [8 6 16 11 9 17]
```

The first column of the integer representation of the proof program (Table 27) is the statement label (row number of the proof program). The second column contains the integer labels of the program names. Columns 3,4 and 5 are the integer labels of the elements of the I/O lists (here we are using nx=2 and ny=1). Null variables are set to 0.

Columns 6 is the integer label of the stored IEP that was used to infer the program of that row and the remaining columns are the elements of the connection list containing the statement labels (row numbers) that formed the sublist of the proof program that is I/O equivalent to the premise of the IEP labeled in column 6. The trailing zeros beyond column 12 have been removed in Table 27.

*Table 27: Integer representation of the proof program of theorem* `thm2` *of the application* `nat[pname ax mach]`

```
 1   13   16   17    0    0    0    0    0    0    0    0
 2   13   17   18    0    0    0    0    0    0    0    0
 3   10   16    0    0   38    1    0    0    0    0    0
 4   10   17    0    0   38    1    0    0    0    0    0
 5   10   18    0    0   38    2    0    0    0    0    0
 6   14   16    0    1   18    3    0    0    0    0    0
 7   14   17    0    2   18    4    0    0    0    0    0
 8   14   18    0    3   18    5    0    0    0    0    0
 9   15   16    0    4   19    3    0    0    0    0    0
10   15   17    0    5   19    4    0    0    0    0    0
11   15   18    0    6   19    5    0    0    0    0    0
12    9    2    1    0   25    7    6    1    0    0    0
13    9    3    2    0   25    8    7    2    0    0    0
14    9    4    5    0   26   10    9    1    0    0    0
15    9    5    6    0   26   11   10    2    0    0    0
16    9    3    1    0   15   13   12    0    0    0    0
17    9    4    6    0   15   14   15    0    0    0    0
18   13   16   18    0   27    8    6   16   11    9   17
```

## 10.7 Connection list reduction

Connection list reduction is a process that identifies statements in a proof program that are redundant. Consider a proof program p=[p[1] p[2] ... p[n]] with premise [p[1] p[2] ... p[m]], lt[m n]. Each derived statement p[i], i=m+1,m+2,...,n, has a connection list that is preceded by the label of the IEP that was used to infer p[i]. The derived statements in a proof program appear as

i=m+1,m+2,...,n
   i p[i] atl[i] clist[i]

The the connection list, clist[i], is given by

clist[i] = [l[i][1] l[i][2] ... l[i][k[i]]]

where k[i] is the length of the premise program of the stored IEP labeled atl[i] and le[l[i][j] i-1], j=1,2,...,k[i], are the row numbers of the sublist of p,

[p[l[i][1]] p[l[i][2]] ... p[l[i][k]]]

that is I/O equivalent to premise of the IEP labeled atl[i] that was employed in the derivation of the statement p[i].

We wish to trace the dependence of the connection list of the conclusion, clist[n], back to the statements of the premise. We call this procedure as *connection list reduction* and is outlined in the following algorithm.

*Algorithm 6: Connection list reduction*

```
b:=order[clist[n]]
do i=n-1 to m+1 [-1]
    if in[i b] then
        b:=minus[b [i]]
        b:=unique[conclst[b clist[i]]]
        b:=order[b]
    endif
enddo
```

The function order[:] places the entries of a list in increasing numerical order. After the do loop is completed the list b should contain only the statement labels of the premise program. If b contains statement labels that are not in the list of labels of the premise then those statements are redundant to the proof of the theorem.

Premise labels that are not contained in the final output of b are also redundant. The presence of redundant statements in the premise indicates that the theorem has been ill defined.

<u>Example</u> :: A step by step reduction of the connection list of the proof of theorem thm2 is shown in Table 28. The first list is the (ordered) connection list of the final statement in the proof program. In this example the last list contains only the premise statement labels indicating that there are no redundant statements in the proof.

*Table 28: Connection list reduction of theorem* thm2

```
[6 8 9 11 16 17]
[6 8 9 11 14 15 16]
[6 8 9 11 12 13 14 15]
[2 6 8 9 10 11 12 13 14]
[1 2 6 8 9 10 11 12 13]
[1 2 6 7 8 9 10 11 12]
[1 2 6 7 8 9 10 11]
[1 2 5 6 7 8 9 10]
[1 2 4 5 6 7 8 9]
[1 2 3 4 5 6 7 8]
[1 2 3 4 5 6 7]
[1 2 3 4 5 6]
[1 2 3 4 5]
[1 2 3 4]
[1 2 3]
[1 2]
```

# 11  Derivable rules of PECR

## 11.1      Some basic derivable rules of PECR

The axioms of PECR are presented as higher order IEPs. Here we will employ PECR as a self referencing tool to explore some properties of PECR itself. All of the proofs shown here were generated by RA's automated theorem prover and validated by VPC's proof checking component. They are presented in the output format used by VPC.

There are several rules of PECR that at first glance appear to be axioms but turn out to be derivable. We start with the most basic rules. The proofs of theorems thm1-thm9 below make use of only the main core axioms per, cr1-cr7, flse1-flse2, the IOT axiom and the substitution rule with the exception of thm8 that employs in addition the supplemental axiom eope.

Program equivalence is reflexive, symmetric and transitive. Theorems thm1 and thm2 establish reflexivity and symmetry, respectively.

thm1

```
typep [a] []
--------------
equiv [a a] []
```

proof

```
  1  typep [a] []
  2  sub [a a] []                    cr4a [1]
  3  equiv [a a] []                  cr4b [2 2]
```

thm2

```
equiv [a b] []
--------------
equiv [b a] []
```

proof

```
  1  equiv [a b] []
  2  typep [a] []                    iot [1]
  3  equiv [a a] []                  thm1 [2]
  4  equiv [b a] []                  sr1 [3 1]
```

Transitivity of program equivalence

```
equiv [a b] []
equiv [b c] []
--------------
equiv [a c] []
```

follows immediately from the substitution rule under equivalence.

There is no commutative rule in the form of an existence axiom for program list
concatenation. However there are valuations where the programs [conc [a b] [c]] and
[conc [b a] [d]] are computable. We have the following equivalence rule.

thm3

```
conc [a b] [c]
conc [b a] [d]
--------------
equiv [d c] []
```

proof

```
  1  conc [a b] [c]
  2  conc [b a] [d]
  3  sub [a c] []              cr5a [1]
  4  sub [b d] []              cr5a [2]
  5  sub [b c] []              cr5b [1]
  6  sub [a d] []              cr5b [2]
  7  sub [c d] []              cr5c [1 6 4]
  8  sub [d c] []              cr5c [2 5 3]
  9  equiv [d c] []            cr4b [8 7]
```

The empty program, ep, is a sublist of every program as is demonstrated by the
following theorem.

thm4

```
typep [a] []
--------------
sub [ep a] []
```

proof

```
  1  typep [a] []
  2  conc [a ep] [b]           cr6a [1]
  3  sub [ep b] []             cr5b [2]
  4  equiv [b a] []            cr6c [2]
  5  sub [ep a] []             sr1 [3 4]
```

It is possible to have a PE of the empty program, ep. The following shows that if c
is a PE of the empty program then c is a PE of any program p that does not violate
the I/O dependency conditions with respect to c, i.e. conc[p c] is type prgm.

thm5

```
ext [ep c] []
conc [p c] [s]
--------------
ext [p c] []
```

proof

```
  1  ext [ep c] []
  2  conc [p c] [s]
  3  typep [p] []              iot [2]
  4  sub [ep p] []             thm4 [3]
```

```
  5  aext [p c] []                     per [1 4 2]
  6  ext [p c] []                      cr1 [5]
```

Axioms cr6a and cr6b state the existence of both left and right concatenation of
any program list with the empty program. Despite the absence of a commutative rule
in the form of an existence axiom for program list concatenation only one
equivalence rule, cr6c, is needed since the following is derivable.

thm6

conc [ep a] [b]
---------------
equiv [b a] []

proof

```
  1  conc [ep a] [b]
  2  typep [a] []                      iot [1]
  3  sub [a a] []                      cr4a [2]
  4  sub [a b] []                      cr5b [1]
  5  sub [ep a] []                     thm4 [2]
  6  sub [b a] []                      cr5c [1 5 3]
  7  equiv [b a] []                    cr4b [6 4]
```

The substitution rule under equivalence should not be applied as an axiom to the
programs [ext [a b] []] and [aext [a b] []]. However it can be shown that the
program [ext [a b] []] satisfies the substitution rule as a derivation.

We first check that the substitution rule under equivalence will be satisfied for
the first input element of [ext [a b] []].

thm7

ext [a b] []
equiv [c a] []
--------------
ext [c b] []

proof

```
  1  ext [a b] []
  2  equiv [c a] []
  3  typep [c] []                      iot [2]
  4  conc [a b] [d]                    cr2 [1]
  5  sub [c c] []                      cr4a [3]
  6  equiv [a c] []                    thm2 [2]
  7  conc [c b] [e]                    sr1 [4 6]
  8  sub [a c] []                      sr1 [5 2]
  9  aext [c b] []                     per [1 8 7]
 10  ext [c b] []                      cr1 [9]
```

The substitution rule under equivalence for the second input element of
[ext [a b] []] is useful when dealing with disjunction and conjunctions. Here we
make use of the supplemental equivalence axiom, eope.

```
thm8

ext [a b] []
equiv [c b] []
--------------
ext [a c] []
```

proof

```
  1  ext [a b] []
  2  equiv [c b] []
  3  aext [a c] []                     eope [1 2]
  4  ext [a c] []                      cr1 [3]
```

Similarly, the substitution rule under equivalence should not be applied as an
axiom to the programs [flse [a] []] and [aflse [a] []]. However the program
[flse [a] []] does satisfy the substitution rule under equivalence as a derivation.

```
thm9

flse [a] []
equiv [b a] []
--------------
flse [b] []
```

proof

```
  1  flse [a] []
  2  equiv [b a] []
  3  typep [b] []                      iot [2]
  4  sub [b b] []                      cr4a [3]
  5  sub [a b] []                      sr1 [4 2]
  6  aflse [b] []                      flse1 [5 1]
  7  flse [b] []                       flse2 [6]
```

## 11.2    Exchange, weakening and contraction

From this point on we will explore some similarities that exist between PECR and
the formal systems of proof theory. In Section 4.9 we discussed an analogy that
exists between natural deduction and PEs in PECR. Recall that judgments in natural
deduction take the form

$\Pi \vdash C$

where $C$ is a single formula and

$\Pi = P_1, P_2, \ldots, P_n$

is a sequent of formulas, $P_i$, i=1,2,...,n. The semantics of the expression $\Pi \vdash C$
asserts that whenever all of the formulas $P_1, P_2, \ldots, P_n$ are true then $C$ is true.

In our formal system, PECR, APs take the place of single formulas and a program
list p=[p[1] p[2] ... p[n]] is used instead of a sequent $\Pi = P_1, P_2, \ldots, P_n$. The statement of entailment, $\Pi \vdash C$, is then replaced by the statement type[c ext[p]] for some AP c.

Some of the analogies that follow will be based on comparisons between the construction rules of PECR and the rules of the sequent calculus. We shall see that analogies of many of the axioms of the sequent calculus can be derived in PECR.

It should be noted that the sequent calculus comes in the two main versions of classical and intuitionistic logic. For intuitionistic logic only a single formula can appear on the right hand side of the turnstile, $\vdash$, whereas for classical logic the right hand side may include sequents. In order to maintain a closer relevance to PECR we will avoid statements that are exclusive to classical logic and where appropriate present only the intuitionistic version.

The sequent calculus has extensions that include rules that involve quantifiers of formulas dependent on variables. These rules will not be considered here because they represent the most significant departure of the sequent calculus from the construction rules of PECR.

From this point on we will use the upper case Greek letters

$\Pi \; \Delta$

to represent sequents and the upper case Latin letters

$A \; B \; C$

to represent single formulas.

<u>Identity</u>

The first axiom of the sequent calculus is the identity axiom

-----
$A \vdash A$

In PECR we can express an identity rule in the form of the combined statements presented in Section 5.3. However these are not axioms in PECR because they follow from the substitution rule.

<u>Exchange</u>

An important property of the sequent calculus is that the order of the formulas in a sequent can be changed. This can be stated in the form of the exchange rule

$\Pi,A,B,\Delta \vdash C$
---------
$\Pi,B,A,\Delta \vdash C$

An exchange rule in PECR can be derived. Since we are dealing with functional programs the reordering of subprograms in a program list is only allowed if the I/O dependency conditions are not violated. Consequently there is no commutative rule for program list concatenations in the form of an existence axiom. However we could have valuations such that conc[a b] and conc[b a] are both type prgm. Suppose that for some programs a, b, p, and d the program u=conc[conc[p conc[a b]] d] has a PE
c, i.e. type[c ext[u]]. We want to show that type[c ext[v]], where
v=conc[conc[p conc[b a]] d].

thm10

```
conc [a b] [e]
conc [p e] [f]
conc [f d] [u]
ext [u c] []
conc [b a] [g]
conc [p g] [h]
conc [h d] [v]
--------------
ext [v c] []
```

proof

```
 1  conc [a b] [e]
 2  conc [p e] [f]
 3  conc [f d] [u]
 4  ext [u c] []
 5  conc [b a] [g]
 6  conc [p g] [h]
 7  conc [h d] [v]
 8  equiv [g e] []              thm3 [1 5]
 9  equiv [f h] []              sr2 [6 8 2]
10  equiv [v u] []              sr2 [3 9 7]
11  ext [v c] []                thm7 [4 10]
```

Here the programs a and b need not be atomic. In this way we have, by analogy,
generalized the exchange rule of the sequent calculus by replacing the single
formulas $A$ and $B$ with sequents. To strictly adhere to the exchange rule in the form
given above we could include in the premises the statements [typeap [a] []] and
[typeap [b] []]. However RA and VPC identify these statements as superfluous.
Throughout the remainder, when making comparisons with the sequent calculus we can
sometimes assume this type of generalization, i.e. we can use program lists where
the axioms of the sequent calculus require single formulas. This follows from the
fact that program lists can be type prgm[1], i.e. atomic programs.

<u>Weakening</u>

The weakening rule in the sequent calculus can be expressed as

$\Pi \vdash C$
-------
$\Pi, A \vdash C$

Here $A$ can be any formula.

We can derive the following rule where the program p of an EP conc[p c],
type[c ext[p]], is weakened by the program list concatenation r=conc[p a] for some
program a. We want to show that type[c ext[r]].

In PECR there is a restriction because we cannot use any program a. In order that
the introduced program, a, does not violate any of the I/O dependency conditions with respect to the existing programs p and c we must include in the premise of the
following theorem the extra conditional statement [conc [r c] [s]].

thm11

```
ext [p c] []
conc [p a] [r]
conc [r c] [s]
--------------
ext [r c] []
```

proof

```
  1  ext [p c] []
  2  conc [p a] [r]
  3  conc [r c] [s]
  4  sub [p r] []              cr5a [2]
  5  aext [r c] []             per [1 4 3]
  6  ext [r c] []              cr1 [5]
```

By repeated use of the exchange rule in the sequent calculus the order of $\Pi, A$ on
the left hand side of the turnstile, $\vdash$ , can be interchanged. Because our analogy
with the exchange rule is subject to restrictions imposed by the I/O dependency
conditions we need to check that weakening in PECR will also work for the program
list concatenation r=conc[a p].

thm12

```
ext [p c] []
conc [a p] [r]
conc [r c] [s]
--------------
ext [r c] []
```

proof

```
  1  ext [p c] []
  2  conc [a p] [r]
  3  conc [r c] [s]
  4  sub [p r] []              cr5b [2]
  5  aext [r c] []             per [1 4 3]
  6  ext [r c] []              cr1 [5]
```

Contraction

In the sequent calculus there is the contraction rule

$$\Pi, A, A \vdash C$$
---------
$$\Pi, A \vdash C$$

In PECR repetitions of programs are allowed if they have an empty output list.
Including a repetition of a program with a nonempty output list is only possible by
introducing new labels for the output list variables (Section 5.3).

Let r=conc[p a]. In order to make some analogy with the above contraction rule we
introduce a second program, a1, such that equiv[a1 a], and construct s=conc[r a1]

=conc[conc[p a] a1]. We want to show that if type[c ext[s]] then it follows that
type[c ext[r]].

thm13

```
conc [p a] [r]
conc [r a1] [s]
ext [s c] []
equiv [a1 a] []
---------------
ext [r c] []
```

proof

```
 1  conc [p a] [r]
 2  conc [r a1] [s]
 3  ext [s c] []
 4  equiv [a1 a] []
 5  typep [r] []             iot [1]
 6  sub [r r] []             cr4a [5]
 7  sub [a r] []             cr5b [1]
 8  conc [r a] [b]           sr1 [2 4]
 9  sub [r b] []             cr5a [8]
10  sub [b r] []             cr5c [8 6 7]
11  equiv [b s] []           sr2 [2 4 8]
12  equiv [r b] []           cr4b [9 10]
13  ext [b c] []             thm7 [3 11]
14  ext [r c] []             thm7 [13 12]
```

## Program extension rule

The program extension rule, per, asserts that if type[c ext[q]] and sub[q p] such
that type[conc[p c] prgm] then the type assignment type[c ext[p]] is valid. We have
accepted that when dealing with program lists in PECR the actions of weakening and
exchange are subject to restrictions. Keeping in mind these restrictions one can
view the program extension rule from a perspective that is suggested by a derivable
result of the sequent calculus.

Let $\Pi(\Delta)$ be a sequent that contains, in any order, all of the formulas that make up
a sequent $\Delta$. Suppose we have

$$\Delta \vdash C$$

By repeated use of the weakening and exchange rules it follows that

$$\Delta \vdash C$$
$$\text{--------}$$
$$\Pi(\Delta) \vdash C$$

By analogy we can interpret the program p of the program extension rule as the end
product of repeated weakening and exchange applied to the program q subject to the
restrictions set by the I/O dependency conditions. From theorems thm10-thm12 it
follows that if p is obtained in this way and type[c ext[q]] then type[c ext[p]].

However this interpretation is misleading in the way the program extension rule is
actually employed when constructing proofs. In PECR we start with a program p that
represents a proof under construction and from it extract a sublist q with a desirable property, namely that type[c ext[q]] for some AP c. After relabeling the
output list elements of c, if necessary, to ensure that type[conc[p c] prgm] we
then apply the program extension rule to perform the type assignment
type[c ext[p]].

## 11.3     Disjunctions

We now include the supplemental axioms eope, disj1-disj6 and cnj1-cnj2 to the main
core axioms of PECR.

In classical logic the statements P∧(A∨B) and (P∨A)∧(P∨B) are logically equivalent.
In PECR there are similar rules that are reflected in the disjunction distribution
rules. Because exchange is subject to constraints in PECR both left and right
distribution rules are required. Theorems thm14-thm15 below generalize the
disjunction distribution rules, dsj3-dsj4.

Theorem thm14 shows that if d=disj[a b], u=conc[conc[p a] q] and
v=conc[conc[p b] q] are type prgm then disj[u v] is type prgm. This combines dsj3a
and dsj4a into a single rule.

thm14

```
disj [a b] [d]
conc [p a] [f]
conc [f q] [u]
conc [p b] [g]
conc [g q] [v]
--------------
disj [u v] [e]
```

proof

```
  1  disj [a b] [d]
  2  conc [p a] [f]
  3  conc [f q] [u]
  4  conc [p b] [g]
  5  conc [g q] [v]
  6  disj [f g] [c]              dsj3a [1 2 4]
  7  disj [u v] [e]              dsj4a [6 3 5]
```

Theorem thm15 shows that if d=disj[a b], s=conc[conc[p d] q], u=conc[conc[p a] q],
v=conc[conc[p b] q] and the disjunction j=disj[u v] are type prgm then equiv[s j].
This combines dsj3b and dsj4b into a single rule.

thm15

```
disj [a b] [d]
conc [p d] [r]
conc [r q] [s]
conc [p a] [f]
conc [f q] [u]
conc [p b] [g]
conc [g q] [v]
disj [u v] [j]
--------------
equiv [s j] []
```

proof

```
 1  disj [a b] [d]
 2  conc [p d] [r]
 3  conc [r q] [s]
 4  conc [p a] [f]
 5  conc [f q] [u]
 6  conc [p b] [g]
 7  conc [g q] [v]
 8  disj [u v] [j]
 9  disj [f g] [c]              dsj3a [1 4 6]
10  equiv [r c] []              dsj3b [1 4 6 2 9]
11  conc [c q] [e]              sr1 [3 10]
12  equiv [e j] []              dsj4b [9 5 7 11 8]
13  equiv [e s] []              sr2 [3 10 11]
14  equiv [s j] []              sr1 [12 13]
```

## Disjunction contraction rule

The disjunction contraction rule, dsj1, can be written in the more general form given by theorem thm16. It states that if d=disj[a b], s=conc[conc[p d] q], u=conc[conc[p a] q] and v=conc[conc[p b] q] are type prgm such that type[c ext[u]] and type[c ext[v]] then the type assignment type[c ext[s]] is valid.

thm16

```
disj [a b] [d]
conc [p d] [r]
conc [r q] [s]
conc [p a] [f]
conc [f q] [u]
conc [p b] [g]
conc [g q] [v]
ext [u c] []
ext [v c] []
--------------
aext [s c] []
```

proof

```
 1  disj [a b] [d]
 2  conc [p d] [r]
 3  conc [r q] [s]
 4  conc [p a] [f]
 5  conc [f q] [u]
 6  conc [p b] [g]
 7  conc [g q] [v]
 8  ext [u c] []
 9  ext [v c] []
10  typeap [c] []              iot [8]
11  typep [c] []               cr7 [10]
12  disj [u v] [e]             thm14 [1 4 5 6 7]
13  aext [e c] []              dsj1 [8 9 12]
14  equiv [c c] []             thm1 [11]
15  equiv [s e] []             thm15 [1 2 3 4 5 6 7 12]
16  ext [e c] []               cr1 [13]
17  ext [s c] []               thm7 [16 15]
```

18  aext [s c] []                    eope [17 14]

Analogies between disjunctions in PECR and disjunctions in formal systems of proof
theory can sometimes become a little stretched. With this in mind there are some
observations that may be useful.

When a disjunction is encountered in a proof program that is under construction we
can split the current proof program into two expanded operand programs, each
associated with an operand of the original disjunction. If there exists a PE that
is common to both of the expanded operand programs then it is also a PE of the
current proof program.

We start with a program s=conc[conc[p d] q], where d=disj[a b]. This is split into
the two expanded operand programs u=conc[conc[p a] q] and v=conc[conc[p b] q].
Suppose that independent derivations for each program, u and v, have been carried
out that lead to a common conclusion, c. We can then apply the generalized
disjunction contraction rule of theorem thm16 to obtain type[c ext[s]].

In the sequent calculus there is the left disjunction rule

$\Pi,A \vdash C \quad \Pi,B \vdash C$
----------------
$\Pi,A \lor B \vdash C$

By repeated use of the weakening rule we can append a sequent, $\Delta$, to the sequents
$\Pi,A$ and $\Pi,B$ to obtain

$\Pi,A,\Delta \vdash C \quad \Pi,B,\Delta \vdash C$
--------------------
$\Pi,A \lor B,\Delta \vdash C$

In this form it is easier to see that the left disjunction rule of the sequent
calculus has some similarity with the generalized disjunction contraction rule,
theorem thm16.

<u>Right disjunction rules</u>

In the sequent calculus there are two right disjunction rules

$\Pi \vdash A$                          $\Pi \vdash B$
--------                          --------
$\Pi \vdash A \lor B$                      $\Pi \vdash A \lor B$

In PECR we have the rule disj6 that is analogous to the first of the right
disjunction rules. In PECR only one rule is needed because we can derive an analogy
of the second of the right disjunction rules as follows.

thm17

ext [p b] []
disj [a b] [c]
conc [p c] [s]
--------------
aext [p c] []

```
proof

  1  ext [p b] []
  2  disj [a b] [c]
  3  conc [p c] [s]
  4  disj [b a] [d]              dsj2a [2]
  5  equiv [c d] []              dsj2b [4 2]
  6  conc [p d] [e]              sr1 [3 5]
  7  aext [p d] []               dsj6 [1 4 6]
  8  ext [p d] []                cr1 [7]
  9  aext [p c] []               eope [8 5]
```

<u>Disjunction contraction rule 2</u>

Theorem thm18 shows that the disjunction contraction rule 2, introduced in Section 6.3, is a theorem. It states that if d=disj[a b] such that type[c ext[a]] and type[b false] then the type assignment type[c ext[d]] is valid.

```
thm18

disj [a b] [d]
ext [a c] []
flse [b] []
--------------
aext [d c] []

proof

  1  disj [a b] [d]
  2  ext [a c] []
  3  flse [b] []
  4  typeap [c] []               iot [2]
  5  typep [c] []                cr7 [4]
  6  equiv [d a] []              dsj5 [1 3]
  7  equiv [c c] []              thm1 [5]
  8  ext [d c] []                thm7 [2 6]
  9  aext [d c] []               eope [8 7]
```

Theorem thm19 generalizes the disjunction contraction rule 2. It states that if d=disj[a b] and s=conc[conc[p d] q], u=conc[conc[p a] q] and v=conc[conc[p b] q] are type prgm such that type[c ext[u]] and type[v false], then the type assignment type[c ext[s]] is valid.

```
thm19

disj [a b] [d]
conc [p d] [r]
conc [r q] [s]
conc [p a] [f]
conc [f q] [u]
conc [p b] [g]
conc [g q] [v]
ext [u c] []
flse [v] []
--------------
aext [s c] []
```

proof

```
 1  disj [a b] [d]
 2  conc [p d] [r]
 3  conc [r q] [s]
 4  conc [p a] [f]
 5  conc [f q] [u]
 6  conc [p b] [g]
 7  conc [g q] [v]
 8  ext [u c] []
 9  flse [v] []
10  typeap [c] []                   iot [8]
11  typep [c] []                    cr7 [10]
12  disj [u v] [e]                  thm14 [1 4 5 6 7]
13  equiv [e u] []                  dsj5 [12 9]
14  equiv [c c] []                  thm1 [11]
15  equiv [s e] []                  thm15 [1 2 3 4 5 6 7 12]
16  equiv [s u] []                  sr1 [15 13]
17  ext [s c] []                    thm7 [8 16]
18  aext [s c] []                   eope [17 14]
```

Disjunction contraction rule 3
------------------------------

The following shows that the disjunction contraction rule 3, introduced in Section 6.3, is a theorem. Theorem thm20 states that if d=disj[a b] is type prgm such that both a and b are type false then the type assignment type[d false] is valid.

thm20

```
disj [a b] [d]
flse [a] []
flse [b] []
--------------
aflse [d] []
```

proof

```
 1  disj [a b] [d]
 2  flse [a] []
 3  flse [b] []
 4  typeap [d] []                   iot [1]
 5  typep [d] []                    cr7 [4]
 6  equiv [d a] []                  dsj5 [1 3]
 7  sub [d d] []                    cr4a [5]
 8  flse [d] []                     thm9 [2 6]
 9  aflse [d] []                    flse1 [7 8]
```

Theorem thm21 generalizes the disjunction contraction rule 3. It states that if d=disj[a b], s=conc[conc[p d] q], u=conc[conc[p a] q] and v=conc[conc[p b] q] are type prgm such that u and v are type false then the type assignment type[s false] is valid.

```
thm21

disj [a b] [d]
conc [p d] [r]
conc [r q] [s]
conc [p a] [f]
conc [f q] [u]
conc [p b] [g]
conc [g q] [v]
flse [u] []
flse [v] []
--------------
aflse [s] []

proof

  1  disj [a b] [d]
  2  conc [p d] [r]
  3  conc [r q] [s]
  4  conc [p a] [f]
  5  conc [f q] [u]
  6  conc [p b] [g]
  7  conc [g q] [v]
  8  flse [u] []
  9  flse [v] []
 10  typep [s] []            iot [3]
 11  sub [s s] []            cr4a [10]
 12  disj [u v] [c]          thm14 [1 4 5 6 7]
 13  equiv [s c] []          thm15 [1 2 3 4 5 6 7 12]
 14  aflse [c] []            thm20 [12 8 9]
 15  flse [c] []             flse2 [14]
 16  sub [c s] []            sr1 [11 13]
 17  aflse [s] []            flse1 [16 15]
```

## 11.4    Conjunctions

Like program list concatenations there is no commutative rule in the form of an existence axiom for conjunctions. However there may be valuations where the programs [conj [a b] [c]] and [conj [b a] [d]] are both computable. We have the following equivalence rule.

```
thm22

conj [a b] [c]
conj [b a] [d]
--------------
equiv [d c] []

proof

  1  conj [a b] [c]
  2  conj [b a] [d]
  3  conc [a b] [e]          cnj2a [1]
  4  conc [b a] [f]          cnj2a [2]
  5  equiv [c e] []          cnj2c [3 1]
  6  equiv [d f] []          cnj2c [4 2]
  7  equiv [f e] []          thm3 [3 4]
```

```
 8  equiv [e c] []                    thm2 [5]
 9  equiv [d e] []                    sr1 [6 7]
10  equiv [d c] []                    sr1 [9 8]
```

In the sequent calculus the left conjunction rules can be expressed as

$$A, \Pi \vdash C$$
$$\text{---------}$$
$$A \wedge B, \Pi \vdash C$$

$$B, \Pi \vdash C$$
$$\text{---------}$$
$$A \wedge B, \Pi \vdash C$$

In PECR analogies of these rules can be derived.

The analogy of the first of the left conjunction rule is derived by the following theorem.

thm23

```
conc [a p] [r]
ext [r c] []
conj [a b] [s]
conc [s p] [t]
conc [t c] [u]
--------------
ext [t c] []
```

proof

```
 1  conc [a p] [r]
 2  ext [r c] []
 3  conj [a b] [s]
 4  conc [s p] [t]
 5  conc [t c] [u]
 6  sub [s t] []                      cr5a [4]
 7  sub [p t] []                      cr5b [4]
 8  conc [a b] [d]                    cnj2a [3]
 9  sub [a d] []                      cr5a [8]
10  equiv [s d] []                    cnj2c [8 3]
11  sub [d t] []                      sr1 [6 10]
12  sub [a t] []                      cr4c [9 11]
13  sub [r t] []                      cr5c [1 12 7]
14  aext [t c] []                     per [2 13 5]
15  ext [t c] []                      cr1 [14]
```

Note that due to the presence of the conjunction program [conj[a b] [c]] the premise is computable only if the programs a and b are type atm.

Because exchange in PECR is subject to constraints we need to check that the first of the analogy of the left conjunction rule will work for the program list concatenation [conc [p a] [r]].

```
thm24

conc [p a] [r]
ext [r c] []
conj [a b] [s]
conc [p s] [t]
conc [t c] [u]
--------------
ext [t c] []

proof

  1  conc [p a] [r]
  2  ext [r c] []
  3  conj [a b] [s]
  4  conc [p s] [t]
  5  conc [t c] [u]
  6  sub [p t] []                    cr5a [4]
  7  sub [s t] []                    cr5b [4]
  8  conc [a b] [d]                  cnj2a [3]
  9  sub [a d] []                    cr5a [8]
 10  equiv [s d] []                  cnj2c [8 3]
 11  sub [d t] []                    sr1 [7 10]
 12  sub [a t] []                    cr4c [9 11]
 13  sub [r t] []                    cr5c [1 6 12]
 14  aext [t c] []                   per [2 13 5]
 15  ext [t c] []                    cr1 [14]
```

The analogy of the second of the left conjunction rule is derived by the following theorem.

```
thm25

conc [b p] [r]
ext [r c] []
conj [a b] [s]
conc [s p] [t]
conc [t c] [u]
--------------
ext [t c] []

proof

  1  conc [b p] [r]
  2  ext [r c] []
  3  conj [a b] [s]
  4  conc [s p] [t]
  5  conc [t c] [u]
  6  sub [s t] []                    cr5a [4]
  7  sub [p t] []                    cr5b [4]
  8  conc [a b] [d]                  cnj2a [3]
  9  sub [b d] []                    cr5b [8]
 10  equiv [s d] []                  cnj2c [8 3]
 11  sub [d t] []                    sr1 [6 10]
 12  sub [b t] []                    cr4c [9 11]
 13  sub [r t] []                    cr5c [1 12 7]
 14  aext [t c] []                   per [2 13 5]
```

```
15  ext [t c] []                        cr1 [14]
```

The following checks that the analogy of the second of the left conjunction rule will also work for the program list concatenation [conc [p b] [r]].

```
thm26

conc [p b] [r]
ext [r c] []
conj [a b] [s]
conc [p s] [t]
conc [t c] [u]
--------------
ext [t c] []

proof

 1  conc [p b] [r]
 2  ext [r c] []
 3  conj [a b] [s]
 4  conc [p s] [t]
 5  conc [t c] [u]
 6  sub [p t] []                        cr5a [4]
 7  sub [s t] []                        cr5b [4]
 8  conc [a b] [d]                      cnj2a [3]
 9  sub [b d] []                        cr5b [8]
10  equiv [s d] []                      cnj2c [8 3]
11  sub [d t] []                        sr1 [7 10]
12  sub [b t] []                        cr4c [9 11]
13  sub [r t] []                        cr5c [1 6 12]
14  aext [t c] []                       per [2 13 5]
15  ext [t c] []                        cr1 [14]
```

In the sequent calculus there is also the right conjunction rule

$$\Pi \vdash A \quad \Pi \vdash B$$
-----------
$$\Pi \vdash A \wedge B$$

In PECR this is expressed in the form of the conjunction axiom cnj1.

## 11.5    Negation, implication and cut rules

There are three rules of the sequent calculus that have not been mentioned so far. These are the negation, implication and cut rules.

<u>Negation</u>

The rules of negation of the sequent calculus can be written in the form

$$\Pi \vdash A$$
-------
$$\neg A, \Pi \vdash$$

$$A, \Pi \vdash$$
-------
$$\Pi \vdash \neg A$$

In PECR most APs do not come with negations so general negation rules are not essential. However if one wishes to introduce a negation of an AP in an application the appropriate negation rules can be included in the list of axioms of the application on an individual basis. (This is only necessary when using VPC. RA has the capacity to find these rules under its automated conjecturing component.)

Implication

The implication rules include formulas involving the implication connective, →.

$\Pi \vdash A \qquad B \vdash C$            $A,\Pi \vdash B$
---------------            ---------
$A{\to}B,\Pi \vdash C$                 $\Pi \vdash A{\to}B$

Analogies of implications in PECR are too stretched to be of any use, mainly because of the range of interpretations that can be attached to an implication.

Note that the formula $A{\to}B$ is logically equivalent to the formula $\neg A \lor B$. The presence of a negation demonstrates the difficulty in associating, by analogy, implications with any formal statements presented as programs. This is why it is better to focus on the stronger analogy between a program extension and a statement of entailment.

Cut rule

The cut rule can be expressed as

$\Pi \vdash A \qquad A,\Delta \vdash B$
---------------
$\Pi,\Delta \vdash B$

In applications of the sequent calculus the cut rule can be useful in shortening the length of proofs. However for any sequent that has a proof that employs the cut-rule there exists a cut-free proof.

In PECR a cut rule is not essential for proof constructions. However it is worth noting that checking for redundant statements in a proof relies on the elimination of intermediate statements of the proof. This is achieved by the connection list reduction algorithm of Section 10.7. The process of connection list reduction can be thought of as an iterative application of a cut rule.

# References

1. J. Hoffman and C. Johnson, Irreversibility in Reversible Systems II: The Incompressible Euler Equations, Chalmers Finite Element Center, Chalmers University of Technology, Göteborg Sweden 2005.

2. K. Zuse, Rechnender Raum. Braunschweig: Friedrich Vieweg and Sohn, 1969.

3. K. Zuse, "Calculating Space", MIT Technical Translation AZT-70-164-GEMIT, Massachusetts Institute of Technology (Project MAC), Cambridge, Mass. 02139. 1970.

4. E. T. Jaynes, Information Theory and Statistical Mechanics, Phys. Rev., 106, 620, 1957.

5. E. T. Jaynes, Information Theory and Statistical Mechanics II, Phys. Rev., 108, 171, 1957.

6. H. Barengregt, W. Dekkers and R. Statman, Lambda Calculus With Types, Perspectives in Logic, Cambridge University Press, 2010.

7. S. R. Buss, Introduction to proof theory, Chapter 1, Handbook of Proof Theory, Eds. S. R. Buss, Elsivier Science, 1998.

8. C. Hall and J. O'Donnell, Discrete Mathematics Using a Computer, Springer-Verlag London, 2000.

9. G. Pantelis, Programs as the Language of Science, ArXiv:1811.05116v2. 2020.

10. S. Wolfram, A New Kind of Science, Wolfram Media, 2002.

11. D. Zeilberger, "Real" analysis is a degenerate case of discrete analysis, transcript of planery talk at ICDEA 2001, Augsburg, Germany, Aug., 2001.

12. M. Petkovsek, H. S. Wilf and D. Zeilberger, A=B, A. K. Peters/CRC Press. 1996.

13. G. Chaitin, Meta Math!: The Quest for Omega, Pantheon Books 2005.

14. G. Chaitin, Information, Randomness and Incompleteness, World Scientific, 1987.

15. G. Chaitin, Algorithmic Information Theory, Cambridge University Press, 1987.

16. G. Chaitin, Information-theoretic Incompleteness, World Scientific, 1992.

17. G. Chaitin, On the Length of Programs for Computing Finite Binary Sequences, Journal of the ACM, 13 (4): 547–569, 1996.

18. G. Chaitin, The Limits of Mathematics, Springer-Verlag, 1998.

19. G. Chaitin, The Unknowable, Springer-Verlag, 1999.

20. G. Chaitin, From Philosophy to Program Size, Tallinn Cybernetics Institute, 2003.

21. G. Chaitin, Thinking about Godel and Turing, World Scientific, 2007.

22. G. Chaitin, Algorithmic information theory some recollections, ArXiv:math/0701164v2, 2007.

23. N. J. Wildberger, Divine Proportions: Rational Trigonometry to Universal Geometry, Wild Egg Books, http://wildegg.com, Sydney, 2005.

24. N. J. Wildberger, One dimensional metrical geometry, Geometriae Dedicata, 128, no.1, (2007), 145-166.

25. N. J. Wildberger, A Rational Approach to Trigonometry, Math Horizons, Nov. 2007, 16-20.

26. G. Japaridze, Introduction to computability logic, Annals of Pure and Applied Logic, 123, 1-99, 2003.

27. G. Pantelis, Program Verification of Numerical Computation, ArXiv:1401.1290v1, 2014.

28. G. Pantelis, Program Verification of Numerical Computation - Part 2, ArXiv:1406.2079v4, 2014.

29. G. Pantelis, A Formal System : Rigourous Constructions of Computer Models, ArXiv:1510.04469v3, 2017.

30. R. Penrose, The Emporer's New Mind, Oxford University Press, 1989.